\documentclass[preprint,12pt]{elsarticle}

\usepackage{amssymb}
\usepackage{mathtools}
\usepackage{enumitem}
\usepackage{tikz}

\usepackage{xcolor} 

\graphicspath{{../figures/}}
\DeclareGraphicsExtensions{.pdf} 

\usepackage{tikz}       
\usepackage{pgfplots}   
\pgfplotsset{compat=1.15} 

\usepackage{amsmath}
\usepackage{amsthm}
\usepackage{amsfonts}
\usepackage{amssymb}

\usepackage{algorithm}
\usepackage{algpseudocode}

\usepackage{array} 

\usepackage[caption=false,font=footnotesize]{subfig}

\usepackage{stfloats} 

\usepackage{url} 

\makeatletter
    \@ifpackageloaded{xcolor}{}{\usepackage{xcolor}}
\makeatother
\makeatletter
    \@ifpackageloaded{needspace}{}{\usepackage{needspace}}
\makeatother

\newcommand{\uppercaseGreek}[1]{%
    \begingroup
    \ucmathlist\MakeUppercase{#1}%
    \endgroup
}

\newcommand{\ucmathlist}{%
    \def\alpha{\mathrm{A}}%
    \def\beta{\mathrm{B}}%
    \let\gamma=\Gamma
    \let\delta=\Delta
    \def\epsilon{\mathrm{E}}%
    \def\varepsilon{\mathrm{E}}%
    \def\zeta{\mathrm{Z}}%
    \def\eta{\mathrm{H}}%
    \let\theta=\Theta
    \let\vartheta=\Theta
    \def\iota{\mathrm{I}}%
    \def\kappa{\mathrm{K}}%
    \let\lambda=\Lambda
    \def\mu{\mathrm{M}}%
    \def\nu{\mathrm{N}}%
    \let\xi=\Xi
    \let\pi=\Pi
    \let\varpi=\Pi
    \def\rho{\mathrm{P}}%
    \def\varrho{\mathrm{P}}%
    \let\sigma=\Sigma
    \def\tau{\mathrm{T}}%
    \let\upsilon=\Upsilon
    \let\phi=\Phi
    \let\varphi=\Phi
    \def\chi{\mathrm{X}}%
    \let\psi=\Psi
    \let\omega=\Omega
}

\makeatletter
    \@ifpackageloaded{amsthm}{}{
        \let\proof\relax
        \let\endproof\relax
        \usepackage{amsthm}
    }
\makeatother
\theoremstyle{plain}

\theoremstyle{definition}
    
    \newtheorem{remark}{Remark}

\makeatletter
\def\renewtheorem#1{%
    \expandafter\let\csname#1\endcsname\relax
    \expandafter\let\csname c@#1\endcsname\relax
    \gdef\renewtheorem@envname{#1}
    \renewtheorem@secpar
}
\def\renewtheorem@secpar{\@ifnextchar[{\renewtheorem@numberedlike}{\renewtheorem@nonumberedlike}}
\def\renewtheorem@numberedlike[#1]#2{\newtheorem{\renewtheorem@envname}[#1]{#2}}
\def\renewtheorem@nonumberedlike#1{
    \def\renewtheorem@caption{#1}
    \edef\renewtheorem@nowithin{\noexpand\newtheorem{\renewtheorem@envname}{\renewtheorem@caption}}
    \renewtheorem@thirdpar
}
\def\renewtheorem@thirdpar{\@ifnextchar[{\renewtheorem@within}{\renewtheorem@nowithin}}
\def\renewtheorem@within[#1]{\renewtheorem@nowithin[#1]}
\makeatother
\input{symbolDef.sty}
\input{mySymbol.sty}
\def\myfactor{0.2}

\begin{document}

\begin{frontmatter}



\title{Topological Signal Processing and Learning:\\  Recent Advances and Future Challenges}

\author[inst1]{Elvin Isufi}

\affiliation[inst1]{organization={Faculty of Electrical Engineering Mathematics and computer Science},
            addressline={Delft University of Technology}, 
            city={Delft},
            country={The Netherlands}}

\author[inst1]{Geert Leus}

\author[inst2]{Baltasar Beferull-Lozano}

\author[inst3]{Sergio Barbarossa}

\author[inst3]{Paolo Di Lorenzo}

\affiliation[inst2]{organization={SIGIPRO, Department, Simula Metropolitan Center for Digital Engineering},
            city={Oslo},
            country={Norway}}

\affiliation[inst3]{organization={Department of Information Engineering, Electronics, and Telecommunications},
            addressline={Sapienza University of Rome}, 
            city={Rome},
            country={Italy}}

\begin{abstract}
%

Developing methods to process irregularly structured data is crucial in applications like gene-regulatory, brain, power, and socioeconomic networks. Graphs have been the go-to algebraic tool for modeling the structure via nodes and edges capturing their interactions, leading  to the establishment of the fields of graph signal processing (GSP) and graph machine learning (GML). Key graph-aware methods include Fourier transform, filtering, sampling, as well as topology identification and spatiotemporal processing. Although versatile, graphs can model only pairwise dependencies in the data. To this end, topological structures such as simplicial and cell complexes have emerged as algebraic representations for more intricate structure modeling in data-driven systems, fueling the rapid development of novel topological-based processing and learning methods. This paper first presents the core principles of topological signal processing through the Hodge theory, a framework instrumental in propelling the field forward thanks to principled connections with GSP-GML. It then outlines advances in topological signal representation, filtering, and sampling, as well as inferring topological structures from data, processing spatiotemporal topological signals, and connections with topological machine learning. The impact of topological signal processing and learning is finally highlighted in applications dealing with flow data over networks, geometric processing, statistical ranking, biology, and semantic communication.

\end{abstract}

\begin{keyword}
topological signal processing, topological deep learning, topological data analysis, graph signal processing, graph machine learning, network science, Hodge theory
\end{keyword}

\end{frontmatter}


\section{Introduction and Motivation}
\label{sec:introduction}

Many technological, biological, and natural systems exhibit data with inherently irregular structures, as observed in critical infrastructure networks, neuroscience, gene regulatory networks, and social interaction systems \cite{karlebach2008modelling,jackson2008social,richiardi2013machine,newman2018networks}. Such data often defy the assumptions of traditional Euclidean-based signal processing and machine learning techniques, rendering these approaches insufficient for capturing their underlying complexities. Consequently, these irregular dependencies have motivated the development of new perspectives and methodologies capable of accommodating their non-Euclidean nature.
In particular, graphs have emerged as the dominant paradigm for modeling irregular data structures by representing pairwise relationships through nodes and edges. This framework has been pivotal in historical advances, such as graphical modeling \cite{lauritzen1996graphical} and network science \cite{jackson2008social,newman2018networks}, as well as in the more recent fields of graph signal processing (GSP) \cite{ortega2018graph} and graph machine learning (GML) \cite{bronstein2021geometric}. GSP extends classical signal processing principles to graph-structured data, enabling tasks such as filtering \cite{isufi2024graph}, spectral analysis \cite{sandryhaila2014discrete}, and signal sampling \cite{tsitsvero2016signals}, \cite{tanaka2020sampling}. GML, on the other hand, leverages graph structures as {\it relational inductive biases} \cite{battaglia2018relational}, to design deep learning architectures that directly capture relations among the data, thus facilitating powerful predictive models through representation learning techniques like graph neural networks (GNNs) \cite{wu2020comprehensive}. These methods have revolutionized applications in social networks, recommendation systems, and biological networks, showcasing their potential for uncovering intricate patterns in relational data. However, the pairwise modeling of graphs imposes limitations in representing more complex relationships, necessitating a broader paradigm.


While graphs are versatile, they represent a simple example of a topological space, limited to capturing pairwise (or dyadic) relationships between data entities \cite{grady2010discrete}. As data grows increasingly complex —such as in gene regulatory networks, social interactions, and neural activity— the richness of interactions among the constituent elements often exceeds the scope of simple dyadic relationships \cite{lambiotte2019networks,majhi2022dynamics,bick2023higher}. Moreover, graph-based techniques are predominantly designed for signals defined on nodes, making them inadequate for analyzing signals associated with higher-dimensional structures, such as flows along edges or signals defined over groups of nodes. These limitations constrain the applicability of graph signal processing and learning methods in scenarios where dependencies span multiple scales and dimensions. Therefore, to effectively capture these data structures and their intricate interdependencies, it is crucial to move beyond graphs and adopt richer topological representations.


Topological representations, such as hypergraphs, simplicial complexes, and cell complexes, provide powerful frameworks for modeling and analyzing multiway relationships \cite{bick2023higher}. Hypergraphs extend the concept of edges to encompass connections among more than two nodes, while simplicial and cell complexes introduce hierarchical structures that facilitate efficient processing across multiple dimensions \cite{berge1984hypergraphs}. Although hypergraphs are highly flexible, their generality introduces an inherent complexity in terms of algebraic representation, mainly based on tensor models \cite{zhang2019introducing,pena2023t,zhang2020hypergraph,pena2023learning,wang2024t}, which might limit practical applicability. In contrast, simplicial and cell complexes endow the domain with hierarchical structures, enabling numerically efficient processing that can handle signals defined at various levels of higher-order structures \cite{grady2010discrete}.
Additionally, simplicial and cell complexes are equipped with advanced algebraic tools, such as Hodge theory and the Hodge Laplacian, which enable spectral analysis and signal decomposition across different topological levels \cite{lim2020hodge}. One of the fundamental capabilities of algebraic topology is to extract {\it global properties} of the space, i.e. {\it invariants}, starting from {\it local relations}. Graphs, which are a simple case of topological space, can only capture global properties associated to connectivity. Higher order structures, like simplicial or cell complexes, enable the extraction of additional invariants, like the number of holes or cavities associated with covering the embedding space with higher order simplicial or cell complexes. The knowledge of these invariants, also known as {\it Betti numbers}, plays a fundamental role in devising signal processing tools tuned to the space where the signals live. Finally, unlike graphs, higher order structures allow a joint processing of signals defined not only on nodes and edges but also on higher-dimensional entities, such as triangles or tetrahedra, ultimately, paving the way for new opportunities in analyzing and learning from complex, multi-scale data. Recent advances in topological signal processing and learning (TSPL) have harnessed these capabilities, enabling innovative methods for filtering, reconstruction, and representation learning that are grounded in the geometric and topological properties of data.

While recent progress in TSPL has demonstrated its potential, current approaches face significant challenges such as developing methods from first principles, ensuring computational efficiency, and achieving seamless integration with machine learning frameworks. Advances in Hodge theory, topological Fourier transform, convolutional filtering, and topological neural networks, have provided a theoretical foundation for the field. However, many techniques remain fragmented across disciplines, hindering their broader adoption and development. This paper seeks to unify the latest advances in TSPL, offering a comprehensive and easily-accessible overview of the methods, challenges, and opportunities in this rapidly evolving domain. By addressing the limitations of graph-based techniques and emphasizing the promise of topological frameworks, we aim to inspire further research and applications of TSPL methods to address the increasingly complex demands of modern data analysis.

\subsection{Related Works and Paper Position}

This paper serves two main purposes. First, it provides a unified, comprehensive, up-to-date, and accessible resource on topological structures and their algebraic foundations. It explores how these structures serve as a framework for signals with irregular characteristics that cannot be represented by graphs and how they connect to key signal processing concepts through Hodge theory. Second, it seeks to unify recent, scattered works in topological signal processing and topological machine learning by linking them to these foundational concepts. Ultimately, the paper aims to serve as a valuable resource for both first-time and advanced readers, helping them identify key contributions in this rapidly evolving field. To this end, we position this work in relation to advances in related areas, as well as to seminal works in topological signal processing and topological machine learning.

\smallskip
\emph{1) Graph signal processing and graph machine learning:} Advances in these areas primarily address pairwise dependencies between data points represented by graphs and focus on processing or learning representations from node values (commonly referred to as graph signals) \cite{shuman2013emerging, ortega2018graph, isufi2024graph, wu2020comprehensive, xia2021graph}. Consequently, much of this work does not emphasize processing data defined on edges or other higher-order structures. These contributions have established foundational concepts, such as the graph Fourier transform, sampling and reconstruction of graph signals, graph convolutions, and graph neural networks. While some works on message-passing graph neural networks include aggregation rules for edge values, they often do so in a black-box manner, treating edge values as features without providing principled insights into how they should interact with neighboring node or edge signals. Connections to TSPL arise primarily through the Hodge Laplacian, which encodes relational locality between topological signals. For instance, the Hodge decomposition and the spectrum of the Hodge Laplacian form the foundation for a topological Fourier transform, strongly tied to GSP concepts. These tools also play a critical role in developing principled topological neural networks.

\smallskip
\emph{2) Network science:} This area focuses on analyzing complex systems and understanding their behavior through higher-order networks, such as simplicial complexes, cell complexes, and hypergraphs \cite{lambiotte2019networks, battiston2020networks, krishnagopal2021spectral, battiston2021physics, bick2023higher, zhang2023higher}. On the one hand, much of this research emphasizes the structure of higher-order networks rather than the data defined on them. For example, studies explore probabilistic models for their structure, algebraic representations, and the relationship between spectral properties and higher-order interactions. Connections to TSPL arise primarily from the simplicial and cell complex representations of higher-order networks and the use of the eigendecomposition of Hodge Laplacians to represent signals in the spectral domain. On the other hand, many works focus on modeling information diffusion and contagion over higher-order networks \cite{iacopini2019simplicial, battiston2020networks, majhi2022dynamics, malizia2024reconstructing, lin2024higher}. TSPL can complement these efforts by characterizing the evolution of such processes from a Fourier perspective, providing sampling and reconstruction strategies, and offering more powerful data-driven embedding techniques compared to higher-order random walks \cite{schaub2020random}. Since then, numerous studies have expanded on these foundational concepts, formalizing methods for convolutional filtering, sampling, and reconstruction, as well as developing the first techniques to learn topologies from signals and manage spatiotemporal topological signals.


\smallskip
\emph{3) Topological data analysis (TDA):} This direction focuses on extracting meaningful shapes and structures present in the data by analyzing them in high-dimensional or complex spaces \cite{carlsson2009topology, edelsbrunner2010computational, wasserman2018topological, otter2017roadmap}. TDA provides a suite of tools, such as persistent homology, which captures topological invariants like connected components, loops, and voids across different scales, offering a robust way to characterize data geometry \cite{ghrist2008barcodes, edelsbrunner2022computational}.
Many TDA methods align conceptually with TSPL, as both frameworks aim to model and process data through its multi-scale and multi-dimensional relationships. Persistent homology, for example, has been used to compute topological summaries that complement signal processing techniques by providing insights into the global and local structure of data. Furthermore, advances in TDA have inspired new approaches in topological machine learning, where persistent diagrams and topological features are integrated with neural network architectures for tasks like classification, clustering, and generative modeling \cite{hensel2021survey}. The core idea of TDA is to use topology as a lens to uncover patterns and relationships in data that may be invisible to traditional methods. Differently, TSPL considers the topology as the support for the data and uses principled topological tools to derive analysis tools tuned to the properties of the topological space where the data lives.

\smallskip
\emph{4) Hypergraph signal processing:} This field focuses on processing signals defined over general topological spaces represented by hypergraphs. The study in \cite{zhang2019introducing} introduced a tensor-based hypergraph signal processing (HGSP) framework to extend graph signal processing (GSP) to high-order interactions. Key advancements include the definition of the hypergraph Fourier space, analysis of hypergraph Fourier transform spectrum properties, development of hypergraph sampling theory, and the fundamentals of hypergraph filter design. This approach was later extended in \cite{pena2023t,pena2023learning}, leveraging a novel tensor-tensor product algebra, a powerful tool for preserving the intrinsic structures of tensors. Notable applications of these frameworks include 3D point cloud analysis \cite{zhang2020hypergraph} and brain functional connectivity inference \cite{bispo2024emergence}. Furthermore, \cite{wang2024t} introduced hypergraph neural network architectures grounded in the HGSP framework, opening new avenues for high-order data analysis.

\smallskip
\emph{5) Topological signal processing:} The first works on topological signal processing began to mature with \cite{barbarossa2016introduction, barbarossa2020topologicalsimplicial, barbarossa2020topological}. In parallel, the first tools using tensor representations for hypergraph signal processing were introduced in \cite{zhang2019introducing}, extending the principles of GSP to hypergraph signals. More recently, the overview article in \cite{schaub2021signal} provided a unified, tutorial-style perspective on signal processing concepts for simplicial complexes and hypergraphs. Building on these early contributions, numerous studies have since formalized key methods; however, these developments remain scattered across various disciplines. In Section~\ref{subsec:spectral}, we unify these scattered advancements, providing a cohesive perspective.

\smallskip
\emph{6) Topological deep learning:} This direction focuses on developing end-to-end representation learning methods for data defined on topological structures, building on the success of graph neural networks \cite{feng2019hypergraph, antelmi2023survey, papillon2023architectures, papamarkou2024position, sanborn2024beyond, hajij2206topological}. Many of the methods and analyses in these works share foundational principles with topological signal processing. For instance, they leverage Hodge theory and Hodge decomposition to construct principled architectures, utilizing approaches such as convolution \cite{ebli2020simplicial, yang2022simplicialnn} or attention mechanisms \cite{battiloro2024generalized}, among others. The core idea is to use topology as a relational inductive bias to enhance deep learning methods, enabling inductive learning solutions that can generalize across different topologies and address the limitations of graph-based neural networks. In this context, we highlight Hodge theory and related topological signal processing concepts as foundational tools for advancing topological deep learning (TDL). These concepts pave the way for principled methods that complement and extend the message-passing paradigm.

\subsection{Outline of the Paper}
\label{subsec:outline}


This paper is organized as follows. Section~\ref{sec:topological} introduces the key elements of topological representations of data structures and topological signals. It builds upon the concept of signals defined on graphs and graph Laplacians, extending these ideas to topological signals and Hodge Laplacians. Section~\ref{sec:hodge} focuses on the spectral processing of topological signals using Hodge theory. It first revisits the graph Fourier transform and then transitions to the topological Fourier transform, highlighting their similarities as well as key differences, such as the interpretation of topological frequencies. Section~\ref{sec:advances} discusses recent advancements in methods for signal representation, sampling and reconstruction, learning topologies from data, and handling spatiotemporal topological signals. This section connects these methods to the core concepts introduced in Sections~\ref{sec:hodge} and~\ref{sec:advances}, and explores how these techniques can be applied to develop and enhance machine and deep learning methods for topological structures. Section~\ref{sec:applications} presents promising application areas tailored to the methods discussed. Finally, Section~\ref{sec:OpenIssues} concludes the paper and outlines key open issues that warrant further research.
\section{Knowledge Representation over Topological Spaces}
\label{sec:topological}



In this section, we formalize the concept of signals over topological spaces, focusing specifically on signals defined over simplicial and cell complexes, alongside the algebraic characterization of their associated domains. To ease exposition, we first revisit signals on graphs in Section~\ref{sec:Signals_on_graphs}, before extending these concepts to topological domains in Section~\ref{sec:Signals_on_complexes}.

\subsection{Signals on Graphs}
\label{sec:Signals_on_graphs}

Let us consider a graph $\mathcal{G} = (\mathcal{V}, \mathcal{E})$ consisting of a set of $N$ vertices $\mathcal{V} = \{1,2,..., N\}$, along with a set of $E$ edges $\mathcal{E} \subset \mathcal{V} \times \mathcal{V}$. Let us also denote the $N \times N$ adjacency matrix as ${\bf A}$ where the element $(i,j)$ is denoted as $A_{i,j}$, $i,j \in \mathcal{V}$. We assume $A_{i,j}>0$, if there is a link from node $j$ to node $i$, i.e., $(j,i) \in \mathcal{E}$, or $A_{i,j}=0$, otherwise. 
The combinatorial Laplacian matrix for an undirected graph with a symmetric adjacency matrix ${\bf A}$ is defined as $\mathbf{L}_0 = {\rm diag}(\mathbf{1}^\top\mathbf{A})-\mathbf{A}$. 

Alternatively, a graph can be represented by its incidence matrix that encodes the incidence relations between vertices and edges. To define the incidence matrix $\mathbf{B}_1$, even if the original graph is undirected, it is necessary to introduce an orientation of the edges. Then, for each edge we have an arrow and, in the case of a binary relation, the entries of  $\mathbf{B}_1$ are defined as follows:
  \begin{equation} \label{inc_coeff_B1}
  \big[\mathbf{B}_{1} \big]_{i,j}=\left\{\begin{array}{rll}
  0, & \text{if node $i$ is not incident on edge $j$} \\
  1,& \text{if node $i$ is the head of arrow $j$} \\
  -1,& \text{if node $i$ is the tail of arrow $j$} \\
  \end{array}\right.
  \end{equation}
It is easy to check that the combinatorial Laplacian matrix, for undirected graphs, can be written as:
\begin{equation}
    \mathbf{L}_{0}=\mathbf{B}_{1}\mathbf{B}_{1}^\top,\label{L_0}
\end{equation}
and its structure is independent from the orientation chosen for the edges.

A signal $\mathbf{x} = [x_1, \ldots, x_N]^\top$ over a graph $\mathcal{G}$ is a mapping from the vertex set to the set of real numbers, i.e., $\mathbf{x}: \mathcal{V} \rightarrow \mathbb{R}$. Here, entry $x_i$ is the signal value associated to node $i\in\stV$. 
Clearly, this definition can be generalized by associating vector or matrix-type data to nodes. The fundamental assumption in graph signal processing is that the algebraic proximities between nodes encoded in $\mtA$ or $\mtL_0$ translate into proximities between the respective signals \cite{ortega2018graph, shuman2013emerging}. Such a coupling can then be used for processing signal $\vcx$ by relying on neighboring information in a similar way as we process temporal and image signals based on temporal or spatial proximities \cite{isufi2024graph}.

\subsection{Signals on Topological Spaces} \label{sec:Signals_on_complexes}


Consider a finite set of vertices $\mathcal{V}$. A $k$-simplex $\mathcal{H}_{k,i}$ is a subset of $\mathcal{V}$ with cardinality $k+1$. A face of $\mathcal{H}_{k,i}$ is a subset with cardinality $k$ and consequently a $k$-simplex has $k+1$ faces. A coface of $\mathcal{H}_{k,i}$ is a $(k + 1)$-simplex that includes $\mathcal{H}_{k,i}$ as a subset.  Two simplices are called lower neighbors if they share a common face, and upper neighbors if they share a common coface. A simplicial complex $\mathcal{X}_{K}$ of order $K$, is a collection of $k$-simplices $\mathcal{H}_{k,i}$, $k = 0, \ldots, K$ such that, if a simplex $\mathcal{H}_{k,i}$ belongs to $\mathcal{X}_{K}$, then all its subsets $\mathcal{H}_{k-1,i} \subset \mathcal{H}_{k,i}$ also belong to $\mathcal{X}_{K}$  (inclusivity property) \cite{goldberg2002combinatorial}. The  set of $k$-simplices in $\mathcal{X}_{K}$ is denoted by ${\cal D}_{k} := \{\mathcal{H}_{k,i}: \mathcal{H}_{k,i} \in \mathcal{X}_{K}\} $, with its cardinality represented as $|{\cal D}_{k}| = N_k$. 

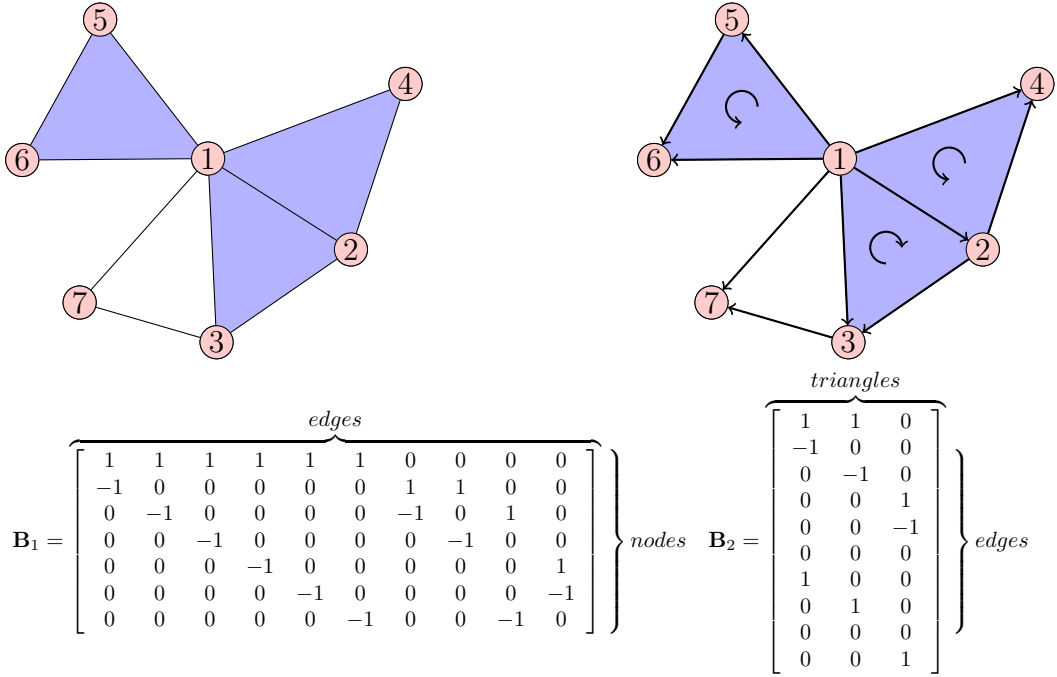
\begin{figure}[t!]
    \centering
    \usetikzlibrary{calc}
\begin{tikzpicture}[scale=0.7]
    \newcommand*\points{653.2278779401073/302.01360907034666/0/1,
                        788.8274758987926/388.3561316827973/1/2,
                        661.260501006842/476.5160082803386/2/3,
                        840.6214214810133/231.08020941549603/3/4,
                        550.7200203762322/169.97868671590825/4/5,
                        476.77803065603206/303.35538555867464/5/6,
                        531.2764234571355/438.7731847144753/6/7}
    \newcommand*\edges{0/1, 0/2, 0/3, 0/4, 0/5, 0/6, 1/2, 1/3, 4/5,6/2}
    \newcommand*\faces{0/1/2/10, 0/1/3/35, 0/4/5/60}
    \newcommand*\scale{0.02}
    \newcommand*\offset{12}

    \foreach \x/\y/\z/\w in \points {
        \node (\z) at (\scale*\x, -\scale*\y) [circle, draw, fill=white, inner sep=1pt] {$\w$};
    }

    \foreach \x/\y/\z/\w in \faces {
        \fill[blue!30] (\x.center) -- (\y.center) -- (\z.center) -- cycle;
    }

    \foreach \x/\y in \edges {
        \draw (\x) -- (\y);
    }

    \foreach \x/\y/\z/\w in \points {
        \node (\z) at (\scale*\x, -\scale*\y) [circle, draw, fill=red!20, inner sep=1pt] {$\w$};
    }

    \foreach \x/\y/\z/\w in \points {
        \node (shift\z) at (\scale*\x + \offset, -\scale*\y) [circle, draw, fill=white, inner sep=1pt] {$\w$};
    }

    \foreach \x/\y/\z/\w in \faces {
        \fill[blue!30] (shift\x.center) -- (shift\y.center) -- (shift\z.center) -- cycle;
    }

    \foreach \x/\y in \edges {
        \draw (shift\x) -- (shift\y);
    }

    \foreach \x/\y in \edges {
        \ifnum\x<\y 
            \draw[->, thick] (shift\x) -- (shift\y); 
        \else
            \draw[->, thick] (shift\y) -- (shift\x); 
        \fi
    }

    \draw[->, thick, black] 
        ($0.5*(shift0.center) + 0.5*(shift5.center) + (0.2,1)$) arc[start angle=0, end angle=270, radius=0.3cm];

    \draw[->, thick, black] 
        ($0.5*(shift0.center) + 0.5*(shift3.center) + (0.5,-0.8)$) arc[start angle=0, end angle=270, radius=0.3cm];
    
    \draw[->, thick, black] 
        ($0.5*(shift0.center) + 0.5*(shift3.center) + (-1,-2.7)$) arc[start angle=270, end angle=0, radius=0.3cm];

    \foreach \x/\y/\z/\w in \points {
        \node (shift\z) at (\scale*\x + \offset, -\scale*\y) [circle, draw, fill=red!20, inner sep=1pt] {$\w$};
    }
\end{tikzpicture}
    \includegraphics[width=\textwidth]{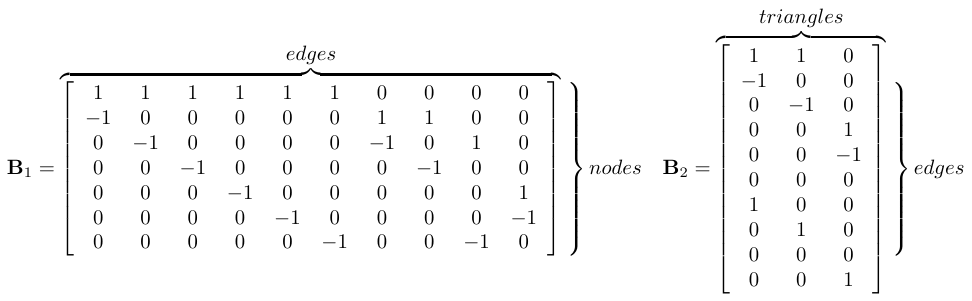}
    \caption{\emph{(top-left)}A simplicial complex of order $K = 2$. It is composed of $7$ nodes, $10$ edges, and $3$ filled triangles (shaded areas). \emph{Example of simplices:} $\stH_{0,1} = \{1\}$, $\stH_{0,2} = \{2\}$, $\stH_{0,3} = \{3\}$; $\stH_{1,1} = \{1,2\}$, $\stH_{1,2} = \{1,3\}$, $\stH_{1,2} = \{2,3\}$; $\stH_{2,1} = \{1,2,3\}$, $\stH_{2,2} = \{1,2,4\}$, $\stH_{2,3} = \{1,5,6\}$. Notice that triangle $\{1,3,7\}$ is empty and not a part of the SC. \emph{(top-right)} The oriented version of the SC following the lexicographi ordering of its vertices. \emph{Example of oriented simplices:} $\stH_{1,1} = [1,2]$, $\stH_{1,2} = [1,3]$, $\stH_{1,2} = [2,3]$; $\stH_{2,1} =[1,2,3]$, $\stH_{2,2} = [1,2,4]$, $\stH_{2,3} = [1,5,6]$. (bottom) Incidence matrices of the oriented simplicial complex.}
    \label{fig:simpComplex}
\end{figure}

In most of the cases the focus is on complexes of order up to two $\mathcal{X}_{2}$, thus having a set of vertices $\mathcal{V}$ with $|\mathcal{V}| = N$, a set of edges $\mathcal{E}$ with $|\mathcal{E}|=E$, and a set of triangles $\mathcal{T}$ with $|\mathcal{T}| = T$, which result in ${\cal D}_{0}={\cal V}$ (simplices of order 0), ${\cal D}_{1}={\cal E}$ (simplices of order 1), and ${\cal D}_{2}={\cal T}$ (simplices of order 2). Figure~\ref{fig:simpComplex} illustrates one such example. Here, edges $\{1,2\}, \{1,3\}, \{2,3\}$ are faces of the filled triangle \{1,2,3\}, and this triangle is a co-face of these edges. Edges $\{1,2\}$ and $\{1,3\}$ are both lower neighbors as they share node 1 (common face), and upper-neighbors as they share triangle \{1,2,3\} (common co-face). Instead, edges $\{1,3\}$ and \{1,7\} are only lower neighbors.

To facilitate computations, it is customary to introduce a reference orientation for simplices of orders $k \ge 1$, i.e., edges, triangles and so on. This reference orientation is a matter of bookkeeping similar to the arbitrary labeling of the nodes in a graph. W.l.o.g., we fix the orientation for a simplex according to the lexicographical ordering of its vertices.  See an example in Figure~\ref{fig:simpComplex}. 

A $k$-simplicial signal is a collection of mappings from the set of all $k$-simplices contained in the complex to real numbers:
\begin{equation}\label{signals}
    \mathbf{x}_k = [x_k(\mathcal{H}_{k,1}),\dots,x_k(\mathcal{H}_{k,i}), \dots, x_k(\mathcal{H}_{k,N_k})]^\top \in \mathbb{R}^{N_k},
\end{equation}
where $x_{k}: {\cal D}_{k} \rightarrow \mathbb{R}$. 
For simplices of order $k \ge 1$, i.e.  edges, triangles, etc., if the signal defined over each simplex represents a flow, then its value is positive if the flow goes in the same direction as the orientation of simplex, or negative in the opposite case.
In general, we define a simplicial complex (SC) signal  as the concatenation  of the signals of each order:
\begin{equation}\label{sc_signal}
    \mathbf{x}_{\mathcal{X}} = \big[\mathbf{x}_0\|\dots\|\mathbf{x}_K\big]\in \mathbb{R}^{\sum_{k=0}^K N_k}.
\end{equation}
For second-order SCs, the $k$-simplicial signals are defined as the following mappings: 
\begin{equation}
    x_{0}: {\cal V} \rightarrow \mathbb{R} , \qquad x_{1}: {\cal E} \rightarrow \mathbb{R} , \qquad x_{2}: {\cal T} \rightarrow \mathbb{R} ,
\end{equation}
representing graph, edge and triangle signals, respectively. In this case, the corresponding SC signal is given by:
\begin{equation}\label{sc_signal_2}
    \mathbf{x}_{\mathcal{X}} = \big[\mathbf{x}_0 \| \mathbf{x}_1 \| \mathbf{x}_2\big] \in \mathbb{R}^{N+E+T}.
\end{equation}

\begin{figure}[t!]
    \centering
    \includegraphics[width=.4\textwidth]{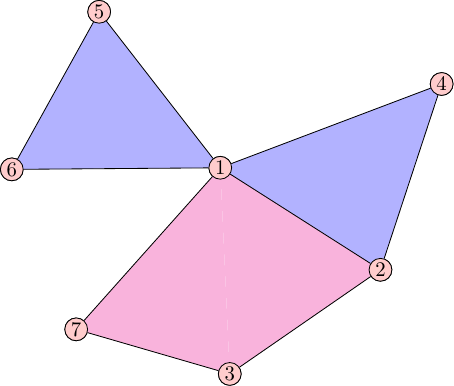}
    \caption{A regular geometric cell complex of order $K = 2$. It is composed of 7 nodes, 9 edges, 2 filled triangles (shaded blue areas), and 1 filled polygon (shaded magenta area). \emph{Examples of 0-cells:} nodes such as \{ \{1\}, \ldots, \{7\} \}. \emph{Examples of 1-cells:} edges such as \{ \{1,2\}, \ldots, \{5,6\} \}. \emph{Examples of 2-cells:} triangles \{ \{1,2,4\}, \{1,5,6\} \} and the quadrilateral \{ \{1,2,3,7\} \}. An orientation can now be set to any cell fro order $k \ge 1$ similar to the simplicial complex. The incidence matrix $\mtB_1$ will capture node-to-edge proximities, while the incidence matrix $\mtB_2$ will capture edge-to-triangle and edge-to-polygon proximities.
    }
    \label{fig:cellcomplex}
\end{figure}

\begin{remark}[Cell complexes.]
These definitions can be generalized to the case of signals defined over cell complexes, which share most of the properties of simplicial complexes, but also some important differences. For instance, differently from simplicial complexes, the inclusivity does not hold \cite{grady2010discrete}, i.e., given a simplex $\mathcal{H}_{k,i}\subset\mathcal{X}_{K}$, not all subsets $\mathcal{H}_{k-1,i} \subset \mathcal{H}_{k,i}$ need to belong to $\mathcal{X}_{k}$. Interestingly, for cell complexes of order 2, this gives rise to the presence of polygon-type relationships among the data. An example is illustrated in Figure~\ref{fig:cellcomplex}, where the presence of the quadrilateral does not imply the presence of triangles or diagonal edges within the quadrlateral itself. For more details on the differences between simplicial and cell complexes refer to \cite{grady2010discrete}, whereas for the role such differences induce in signal representation refer to \cite{sardellitti2021topological, roddenberry2022signal,sardellitti2024topological}. \qed
\end{remark}

\subsection{Algebraic Representation} 

The structure of a simplicial complex ${\cal X}_{K}$  is completely characterized by its set of incidence matrices $\mathbf{B}_{k}$, $k=1, \ldots, K$. Extending the concept from graphs, the entries of the incidence matrix $\mathbf{B}_{k}$ specify which $k$-simplices are incident to which $(k-1)$-simplices. We use the notation $\mathcal{H}_{k-1,i} \sim \mathcal{H}_{k,j}$ to  indicate two simplices with the same orientation,  and  $\mathcal{H}_{k-1,i} \not\sim \mathcal{H}_{k,j}$ to indicate that  they have opposite orientation. The entries of $\mathbf{B}_{k}$ are defined as:
  \begin{equation} \label{inc_coeff}
  \big[\mathbf{B}_{k} \big]_{i,j}=\left\{\begin{array}{rll}
  0, & \text{if} \; \mathcal{H}_{k-1,i} \not\subset \mathcal{H}_{k,j} \\
  1,& \text{if} \; \mathcal{H}_{k-1,i} \subset \mathcal{H}_{k,j} \;  \text{and} \; \mathcal{H}_{k-1,i} \sim \mathcal{H}_{k,j}\\
  -1,& \text{if} \; \mathcal{H}_{k-1,i} \subset \mathcal{H}_{k,j} \;  \text{and} \; \mathcal{H}_{k-1,i} \not\sim \mathcal{H}_{k,j}\\
  \end{array}\right. .
  \end{equation}
As an example, considering a simplicial complex $\mathcal{X}_{2}$ of order two, we have two incidence matrices: the node-to-edge incidence matrix $\mathbf{B}_{1} \in \mathbb{R}^{N \times E}$, and the edge-to-triangle incidence matrix $\mathbf{B}_{2} \in \mathbb{R}^{E \times T}$. The latter are illustrated in Figure~\ref{fig:simpComplex}. From the incidence information, we can build the higher-order Hodge Laplacian matrices, of orders $k=0, \ldots, K$, as:
\begin{align}
&\mathbf{L}_{0}=\mathbf{B}_{1}\mathbf{B}_{1}^\top,\label{Laplacian0}\\
&\mathbf{L}_{k}=\underbrace{\mathbf{B}_k^{\top}\mathbf{B}_{k}}_{\mathbf{L}_k^{(d)}}+\underbrace{\mathbf{B}_{k+1}\mathbf{B}_{k+1}^\top}_{\mathbf{L}_k^{(u)}}, \; k=1, \ldots, K-1, \label{Laplaciank}\\
&\mathbf{L}_{K}=\mathbf{B}_{K}^\top\mathbf{B}_{K}.\label{LaplacianK}
\end{align}
Here, $\mtL_0$ is the combinatorial graph Laplacian defined in Section~\ref{sec:Signals_on_graphs} and captures node-to-node proximities. All Laplacian matrices of intermediate orders $k=1, \ldots, K-1$, contain two terms: The first term $\mathbf{L}^{(d)}_k$, known as  down Laplacian, encodes the lower adjacency of $k$-order simplices; the second term $\mathbf{L}_k^{(u)}$, known as upper Laplacian, encodes the upper adjacency of $k$-order simplices. Thus, for example, two edges are lower adjacent if they share a common vertex, whereas they are upper adjacent if they are faces of a common triangle. 
Note that the vertices of a graph can only be upper adjacent, if they are incident to the same edge. This is why the Laplacian $\mathbf{L}_0$ contains only one term, and it corresponds to the usual graph Laplacian. Similar definitions apply also for the higher-order Hodge Laplacians of cell-complexes, upon defining the proper incidence relationships between the elements (i.e., the cells) of the domain \cite{roddenberry2022signal,sardellitti2024topological}. 

A key property of the Hodge Laplacians is that they make possible to extract some fundamental global properties (invariants) of the complex. More specifically, the dimension of the kernel of the Hodge Laplacian of order $k$ is equal to the Betti number $\beta_k$ of order $k$ and its value represents the number of connected components, for $k=0$, the number of holes, for $k=1$, the number of 3D cavities, for $k=2$, and so on.

{The Hodge Laplacian matrices play the equivalent role for topological signals of the graph Laplacian pays for graph signals. However, depending on the setting, we may be interested in processing solely a simplicial signal (e.g., edge flows) and have no other signal available, or process jointly the SC signal (e.g., joint processing of node, edge, and triangle signals). We shall detail the differences in these two cases in the next two sections.
}

\subsection{Hodge Decomposition}
\label{subsec:hodgeDec}

One useful property of the Hodge Laplacian is their link with the Hodge decomposition. In particular, given the $k$-th intermediate Hodge Laplacian $\mtL_k = \mtB_k^\top\mtB_k + \mtB_{k+1}\mtB_{k+1}^\top$, the Hodge decomposition states that the signal space associated with each simplex of order $k$ can be decomposed as the direct sum of the following three orthogonal subspaces \cite{lim2020hodge}:
\begin{equation}\label{eq.HodgeDec}
    \fdR^{N_k} = \underbrace{\textnormal{span}(\mtB_k^\top)}_{\textnormal{(a) gradient space}}~~\oplus ~~\underbrace{\textnormal{span}(\mtB_{k+1})}_{\textnormal{(b) curl space}}~~\oplus ~~\underbrace{\textnormal{kernel}(\mtL_k)}_{\textnormal{(c) harmonic space}}.
\end{equation}
This implies that the $k$th simplicial signal space is composed of three subspaces, namely \emph{the gradient space} $\textnormal{span}(\mtB_k^\top)$, \emph{the curl space} $\textnormal{span}(\mtB_{k+1})$, and \emph{the harmonic space} $\textnormal{kernel}(\mtL_k)$. In turn, the Hodge decomposition implies that every signal $\vcx_k$ of order $k$ can be decomposed as:
\begin{equation}\label{eq.Hodgesignal}
    \vcx_k = \underbrace{\mtB_k^\top\vctx_{k-1}}_{\textnormal{(a) gradient component}} + \underbrace{\mtB_{k+1}\vctx_{k+1}}_{\textnormal{(b) curl component}} + \underbrace{\vctx_k}_{\textnormal{(c) harmonic component}}.
\end{equation}
That is, there exists three signals $\vctx_{k-1}, \vctx_{k}, \vctx_{k+1}$ of respective orders $k-1$, $k$, $k+1$ that can express the signal. In other words, the decomposition in \eqref{eq.Hodgesignal} shows how the inter-simplex couplings imposed by the Hodge Laplacians translate into inter-signal couplings across different levels. Such couplings yielding from the decompositions in \eqref{eq.HodgeDec} and \eqref{eq.Hodgesignal} carry the following interpretation when discussing edge signals $\vcx_1$ (i.e., $k = 1$) \cite{barbarossa2020topologicalsimplicial, yang2022simplicial}:
\begin{enumerate}[label=(\alph*)]
    \item \emph{Gradient space and gradient component:} The space $\textnormal{span}(\mtB_1^\top)$ is called the \emph{gradient space}. An edge flow signal $\vcx_{\textnormal{1,g}} = \mtB_1^\top\vctx_{0}$ living in this space is referred to as a \emph{gradient flow} and it can be obtained by differentiating node signals $\vctx_{0}$ along the edges connecting them. 
    The component of an edge signal ${\bf x}_1$ living in the gradient space is referred to as the \emph{gradient component} (a.k.a., the irrotational component).
    \item \emph{Curl space and curl component:} The space $\textnormal{span}(\mtB_{2})$ is called the \emph{curl space}. An edge flow signal $\vcx_{1,\textnormal{c}} = \mtB_{2}\vctx_{2}$ living in this space is referred to as a \emph{curl flow} and it can be induced by some triangle signals $\vctx_2$. 
    The component of an edge signal ${\bf x}_1$ living in the curl space is referred to as the \emph{curl component} (a.k.a., the solenoidal component).
    \item \emph{Harmonic space and harmonic component:} The space $\textnormal{kernel}(\mtL_k)$ is called the \emph{harmonic space}. 
    An edge flow signal $\vctx_k$ living in this space is referred to as a \emph{harmonic flow} and it cannot be induced from adjacent simplicial signals. The component of an edge signal ${\bf x}_1$ living in the harmonic space is referred to as the \emph{harmonic component}. 
\end{enumerate}

The Hodge decomposition shows how the topological proximities between different simplices translate into inter-simplicial couplings. While the decomposition holds for any simplicial signal, the terminology is more intuitive when discussing edge flows but often it is used also for a more general setting. We shall see in Section~\ref{subse:tft} how this Hodge decomposition ties with the Fourier analysis of topological signals.

\smallskip
\noindent\textbf{Performance-complexity tradeoff.} Moving beyond graphs introduces additional computational complexity, which must be justified by a corresponding improvement in performance. The advantages of TSP become particularly evident when data exhibit multi-way relationships across various hierarchical levels, which traditional graph-based approaches fail to fully capture. For instance, in the case of edge flows, the Hodge decomposition in \eqref{eq.Hodgesignal} identifies three distinct signal components: gradient, curl, and harmonic. As shown in \eqref{eq.Hodgesignal}, the curl and harmonic components cannot be effectively extracted without incorporating higher-order structures such as triangles or cells. When predictive tasks strongly depend on these components, graph-based methods inevitably lose valuable information, while topological methods  can effectively preserve and capture it, resulting in superior performance in such contexts. This is even more critical when addressing applications where data is defined over second- or higher-order simplicial or cell structures, necessitating the development of more advanced and tailored TSP methods. Finally, the effectiveness of TSP relies heavily on the accurate definition of the topological domain. In cases where such knowledge is absent or incomplete, task-specific topology learning methods must be designed to infer the optimal topological structure that enhances performance while managing complexity. Promoting sparsity in the topological representation should serve as a key principle, as it provides a practical means to strike a balance between computational efficiency and performance gains. By addressing these challenges, TSP has the potential to unlock new opportunities for processing and learning from complex, multi-scale data in diverse applications.


\subsection{Dirac Operator and Dirac Decomposition}
\label{subsec:hodgeDec}

The Hodge Laplacian and Hodge decomposition are conventionally used to represent, analyse, and process signals within a given simplicial order. When signals across different simplicial orders are present, a joint analysis and processing may carry useful information. Since topological complexes rely on the assumption that simplices influence each other only in consecutive orders, then a natural way is to provide a representation that enables topological signal processing across consecutive simplices. The latter is possible via the concept of the Dirac operator which we elaborate next \cite{bianconi2021topological,calmon2023dirac}.

\smallskip
\noindent\textbf{Dirac operator.} Focusing for simplicity on a simplicial complex $\stX_2$ of order $K = 2$, the Dirac operator reads as:
\begin{equation}\label{eq.Dirac}
    \mtD_\stX := \begin{bmatrix}
\vcZeros & \mtB_1 & \vcZeros\\
\mtB_1^\top & \vcZeros & \mtB_2\\
\vcZeros & \mtB_2^\top & \vcZeros
\end{bmatrix} = 
\underbrace{
\begin{bmatrix}
\vcZeros & \mtB_1 & \vcZeros\\
\mtB_1^\top & \vcZeros &\vcZeros \\
\vcZeros & \vcZeros & \vcZeros
\end{bmatrix}}_{\mtD_\stX^{(d)}} + 
\underbrace{\begin{bmatrix}
\vcZeros & \vcZeros & \vcZeros\\
\vcZeros & \vcZeros & \mtB_2\\
\vcZeros & \mtB_2^\top & \vcZeros
\end{bmatrix}}_{\mtD_\stX^{(u)}}
\end{equation}
where we refer to $\mtD_\stX^{(d)}$ and $\mtD_\stX^{(u)}$ as the down and up Dirac operator, respectively. The Dirac operator is such that its square gives a block diagonal concatenation of the Hodge Laplacians, i.e., $\mtD_\stX^2 = \textnormal{blkdiag}(\mtL_0, \mtL_1, \mtL_2)$.

\smallskip
\noindent\textbf{Application to topological complex signals.} When applied to a SC signal, the Dirac operator yields the shifted signal:
%
%
\begin{equation}\label{eq.DiracShift}
    \vcx_\stX^{(1)} =  \begin{bmatrix}
\vcx_0^{(1)}\\
\vcx_1^{(1)}\\
\vcx_2^{(1)}
\end{bmatrix} = \mtD_\stX\vcx_\stX = 
\begin{bmatrix}
\mtB_1\vcx_1\\
\mtB_1^\top\vcx_0 + \mtB_2\vcx_2\\
\mtB_2^\top\vcx_1
\end{bmatrix}.
\end{equation}
The shifted node signal $\vcx_0^{(1)} = \mtB_1\vcx_1$ sums the edge flows flowing into a node and is referred to as the divergence of the edge flow; the shifted edge signal $\vcx_1^{(1)} = \mtB_1^\top\vcx_0 + \mtB_2\vcx_2$ is the linear combination of a gradient flow obtained from the nodes and a curl flow induced by triangle signals; and the shifted triangle signal $\vcx_2^{(1)} = \mtB_2^\top\vcx_1$ is the curl signal of the edge flow. The operation in \eqref{eq.DiracShift} shows how we can translate inter-topological couplings into inter-topological signal operations. The latter has been used in a few recent works to filter topological signals on simplicial complexes in a consistent way \cite{calmon2023dirac}, or design principled topological neural architectures \cite{battiloro2024generalized}.  

\smallskip
\noindent\textbf{Dirac decomposition.} Similar to the Hodge decomposition in \eqref{eq.HodgeDec}, the Dirac decomposition states that the signal space associated with a topological signal $\vcx_\stX$ can also be decomposed as the direct sum of the following three orthogonal subspaces reminiscent of \eqref{eq.Dirac}:
\begin{equation}\label{eq.HodgeDec2}
    \fdR^{\sum_{k = 0}^K N_k} = \underbrace{\textnormal{span}(\mtD_\stX^{(d)})}_{\textnormal{(a) joint gradient space}}~~\oplus ~~\underbrace{\textnormal{span}(\mtD_\stX^{(u)})}_{\textnormal{(b) joint curl space}}~~\oplus ~~\underbrace{\textnormal{kernel}(\mtD_\stX)}_{\textnormal{(b) joint harminic space}}
\end{equation}
where by following the same terminology as for the Hodge decomposition, we refer to (a) $\textnormal{span}(\mtD_\stX^{(d)})$ as \emph{the joint gradient space}; (b) $\textnormal{span}(\mtD_\stX^{(u)})$ as \emph{the joint curl space}; and (c) $\textnormal{kernel}(\mtD_\stX)$ as \emph{the joint harmonic space}. While the interpretation of these subspaces is more involved than that of the Hodge decomposition in \eqref{eq.HodgeDec}, they still allow decomposing a topological signal $\vcx_\stX$ into the sum of a signal $\vctx_\stX^{(d)} \in \textnormal{span}(\mtD_\stX^{(d)})$, a signal $\vctx_\stX^{(u)} \in \textnormal{span}(\mtD_\stX^{(u)})$, and a harmonic signal $\vctx_\stX \in \textnormal{kernel}(\mtD_\stX)$, i.e., 
\begin{equation}\label{eq.Diracsignal}
    \vcx_k = \underbrace{\vctx_{\stX}^{(d)}}_{\textnormal{(a) joint gradient component}} + \underbrace{\vctx_{\stX}^{(u)}}_{\textnormal{(b) joint curl component}} + \underbrace{\vctx_{\stX}}_{\textnormal{(c) joint harmonic component}}.
\end{equation}
We refer the reader to \cite{calmon2023dirac} for more details on the Dirac decomposition of topological signals.

\section{Spectral Processing}
\label{sec:hodge}

In this section, we show how the Hodge and Dirac decomposition can be used for a spectral analysis of topological signals. We first recall the basic concepts of the graph Fourier transform in Section~\ref{subse:gft} and then define the more general topological Fourier transform in Section~\ref{subse:tft}. We conclude by discussing the spectral duality for joint topological complex signals in Section~\ref{subse:dirac}.

\subsection{Graph Fourier Transform}
\label{subse:gft}

Given an undirected graph $\stG = (\stV, \stE)$ with graph Laplacian $\mtL$ and graph signal $\vcx$, the graph Fourier transform (GFT) of signal $\vcx$ is the signal projection onto the Laplacian eigenspace. 
%
More specifically, given the eigendecomposition $\mtL = \mtU\mtLambda\mtU^\Hr$, the GFT of $\vcx$ is $\vchx = \mtU^\Hr\vcx$. The eigenvectors $\mtU = [\vcu_1, \ldots, \vcu_N]$ serve as the spectral basis expansion for the graph signal $\vcx$, and the GFT coefficient $\hat{x}_i$ is the weight indicating how much eigenvector $\vcu_i$ contributes to represent the signal. Following the analogy with the classical Fourier transform \cite{sandryhaila2014discrete}, the eigenvalues in $\mtLambda = \diag(\lambda_1, \ldots,\lambda_N)$ contain the so-called graph frequencies.

The Fourier notion of the projection $\vchx = \mtU^\Hr\vcx$ comes from the fact that we can view each eigenvector $\vcu_i = [u_{i1}, \ldots, u_{iN}]^\top$ as a graph signal and analyze its variability w.r.t.~the graph $\stG$. One way to do this is via the quadratic variation $\QV(\vcu_i) = \vcu_i^\Hr\mtL\vcu_i = \lambda_i$, which indicates how smooth $\vcu_i$ is over the graph $\stG$. Thus, we can sort the eigenvectors based on their variability $0 = \QV(\vcu_1) \le \QV(\vcu_2) \le \ldots \le \QV(\vcu_N) $, which implies an ordering of the graph frequencies $0 = \lambda_1 \le \lambda_2 \le \ldots \le \lambda_N$. As a consequence, we refer to the eigenvalues $\lambda_i$ close to $0$ as low frequencies and to eigenvalues $\lambda_i \gg 0$ as high frequencies. Hence, the GFT coefficient $\hat{x}_i$ indicates how much the eigenvector signal basis $\vcu_i$ contributes to the variability of the graph signal $\vcx$. We shall discuss next that a similar, yet slightly more involved, Fourier analysis can be derived also for topological signals of any order.

\subsection{Topological Fourier Transform}
\label{subse:tft}

As for the graph Laplacian, any Hodge Laplacian of order $k$ enjoys an eigendecomposition of the form
\begin{equation}
    \mtL_k = \mtL_k^{(d)} + \mtL_k^{(u)} =  \mtB_k^\top\mtB_k + \mtB_{k+1}\mtB_{k+1}^\top = \mtU_k\mtLambda_k\mtU_k^\top 
\end{equation}
with orthogonal eigenvector matrix $\mtU_k = [\vcu_{k,1}, \ldots, \vcu_{k,N_k}] \in \mathbb{R}^{N_k \times N_k}$ and eigenvalue matrix $\mtLambda_k = \diag(\lambda_{k,1}, \ldots, \lambda_{k,N_k})$. Then, the \emph{topological Fourier transform} (TFT) of a signal $\vcx_k$ is given by the projection onto the eigenvectors $\mtU_k$, i.e., $\vchx_k = \mtU_k^\top \vcx_k$ \cite{barbarossa2020topologicalsimplicial}. As for the GFT, the $i$-th entry of $\vchx_k$, i.e., $\hat{x}_{k,i}$ represents the weight of eigenvector $\vcu_{k,i}$ in expressing signal $\vcx_k$. The inverse TFT is given by $\vcx_k = \mtU_k \vchx_k$. Notice that the GFT discussed in Section~\ref{subse:gft} is the special case of the TFT for $k = 0$. While in principle the TFT and the GFT are quite similar, they differ substantially in terms of interpretation as we elaborate in the sequel.

\smallskip
\noindent\textbf{Interpreting the TFT.} The Hodge Laplacian eigendecomposition has a correspondence with the Hodge decomposition in \eqref{eq.HodgeDec}. More specifically, from \cite[Proposition 4]{yang2022simplicial} it is possible to rearrange the eigenvectors in $\mtU_k$ and eigenvalues in $\mtLambda_k$ respectively as:
\begin{equation}
    \mtU_k = [\mtU_{k,\textnormal{g}}, \mtU_{k,\textnormal{c}}, \mtU_{k,\textnormal{h}}]~~\textnormal{and}~~\mtLambda_k = \textnormal{blkdiag}(\mtLambda_{k,\textnormal{g}}, \mtLambda_{k,\textnormal{c}}, \mtLambda_{k,\textnormal{h}}).
\end{equation}
Focusing again to edge signals, $k = 1$, we can observe the following:
\begin{enumerate}[label=(\alph*)]
\item \emph{Gradient space and gradient frequencies:} The eigenvectors $\mtU_{1,\textnormal{g}} \in \mathbb{R}^{N_1\times N_{\textnormal{g}}}$ span the gradient space $\textnormal{span}(\mtB_1^\top)$ with dimension $N_{\textnormal{g}}$. Hence, the component $\vchx_{1,\textnormal{g}} = \mtU_{1,\textnormal{g}}^\top\vcx_1$ represents the gradient component of the TFT. The eigenvectors $\vcu_\textnormal{g} \in \mtU_{1,\textnormal{g}}$ also carry a notion of variability via the quadratic variation w.r.t.~the Hodge Laplacian $\mtL_1$, i.e.,
\begin{equation}\label{eq:qvGrad}
    \QV(\vcu_\textnormal{g}) = \vcu_\textnormal{g}^\top\mtL_1\vcu_\textnormal{g} = \|\mtB_1\vcu_\textnormal{g} \|_2^2 + \|\mtB_2^\top\vcu_\textnormal{g} \|_2^2 = \|\mtB_1\vcu_\textnormal{g} \|_2^2 = \lambda_{\textnormal{g}}
\end{equation}
where $\mtB_2^\top\vcu_\textnormal{g} = \mathbf{0}$ since $\vcu_\textnormal{g}$ is a gradient flow. Thus, the eigenvalue $\lambda_{\textnormal{g}}$ is the squared $\ell_2-$norm of the divergence $\mtB_1\vcu_\textnormal{g}$ of the corresponding gradient eigenvector. We can use this quadratic variation to sort the gradient eigenvectors, and correspondingly, the divergence variation of the corresponding eigenvectors in an ascending order $0 < \QV(\vcu_{\textnormal{g},i}) \le \QV(\vcu_{\textnormal{g},j})$ implying an ordering of the eigenvalues $0< \lambda_{\textnormal{g},i} < \lambda_{\textnormal{g},j}$ for $i,j \in \{1, \ldots, N_{\textnormal{g}}\}$. In turn, this ordering carries the same meaning of variability as for the graph frequencies but now it measures the variability of the total divergence. We refer to the eigenvalues $\lambda_{\textnormal{g}}$ associated to the gradient eigenvectors $\mtU_{1,\textnormal{g}}$ as the \emph{gradient frequencies}.
\item \emph{Curl space and curl frequencies:} Analogously, the eigenvectors $\mtU_{1,\textnormal{c}} \in \mathbb{R}^{N_1\times N_{\textnormal{c}}}$ span the curl space $\textnormal{span}(\mtB_2)$ with dimension $N_{\textnormal{c}}$. The projection $\vchx_{1,\textnormal{c}} = \mtU_{1,\textnormal{c}}^\top\vcx_1$ represents the curl component of the TFT, where the eigenvectors $\vcu_\textnormal{c} \in  \mtU_{1,\textnormal{c}}$ do carry a notion of variability that is different from that seen above. Using again the quadratic variation w.r.t. the $\mtL_1$ Hodge Laplacian we have
\begin{equation}\label{eq:qvcurl}
    \QV(\vcu_\textnormal{c}) = \vcu_\textnormal{c}^\top\mtL_1\vcu_\textnormal{c} = \|\mtB_1\vcu_\textnormal{c} \|_2^2 + \|\mtB_2^\top\vcu_\textnormal{c} \|_2^2 = \|\mtB_2^\top\vcu_\textnormal{c} \|_2^2 = \lambda_{\textnormal{c}}
\end{equation}
where $\mtB_1\vcu_\textnormal{c} = \mathbf{0}$ since $\vcu_\textnormal{c}$ is a curl flow. Now, the eigenvalue $\lambda_{\textnormal{c}}$ is the squared $\ell_2-$norm of the total curl $\mtB_2^\top\vcu_\textnormal{c}$ of the corresponding curl eigenvector. Using this interpretation of the quadratic variation, we can sort separately the curl eigenvectors, and correspondingly, the total curl variation of the corresponding eigenvectors in an ascending order $0 < \QV(\vcu_{\textnormal{c},i}) \le \QV(\vcu_{\textnormal{c},j})$ implying an ordering of the curl eigenvalues $0< \lambda_{\textnormal{c},i} < \lambda_{\textnormal{c},j}$ for $i,j \in \{1, \ldots, N_{\textnormal{c}}\}$. We refer to the eigenvalues $\lambda_{\textnormal{c}}$ associated to the curl eigenvectors $\mtU_{1,\textnormal{c}}$ as the \emph{curl frequencies}.
\item \emph{Harmonic space and harmonic frequencies:} Finally, the eigenvectors $\mtU_{1,\textnormal{h}} \in \mathbb{R}^{N_1\times N_{\textnormal{h}}}$ span the harmonic space $\textnormal{kernel}(\mtL_k)$ with dimension $N_{\textnormal{h}}$. The projection $\vchx_{1,\textnormal{h}} = \mtU_{1,\textnormal{h}}^\top\vcx_1$ represents the harmonic component of the TFT, where the eigenvectors $\vcu_\textnormal{h} \in  \mtU_{1\textnormal{h}}$ are all associated to the zero eigenvalue since
\begin{equation}\label{eq:qvharm}
    \QV(\vcu_\textnormal{h}) = \vcu_\textnormal{h}^\top\mtL_1\vcu_\textnormal{h} = \|\mtB_1\vcu_\textnormal{h} \|_2^2 + \|\mtB_2^\top\vcu_\textnormal{h} \|_2^2 = 0 = \lambda_{\textnormal{h}}.
\end{equation}
Consequently, we will refer to the eigenvalues $\lambda_{\textnormal{h}} = 0$ as the \emph{harmonic frequencies}. These harmonic frequencies correspond to a global conservative flow, i.e., a flow signal that does not have any gradient or curl component.
\end{enumerate}

\begin{figure}[t!]
    \centering
    \includegraphics[width=\textwidth]{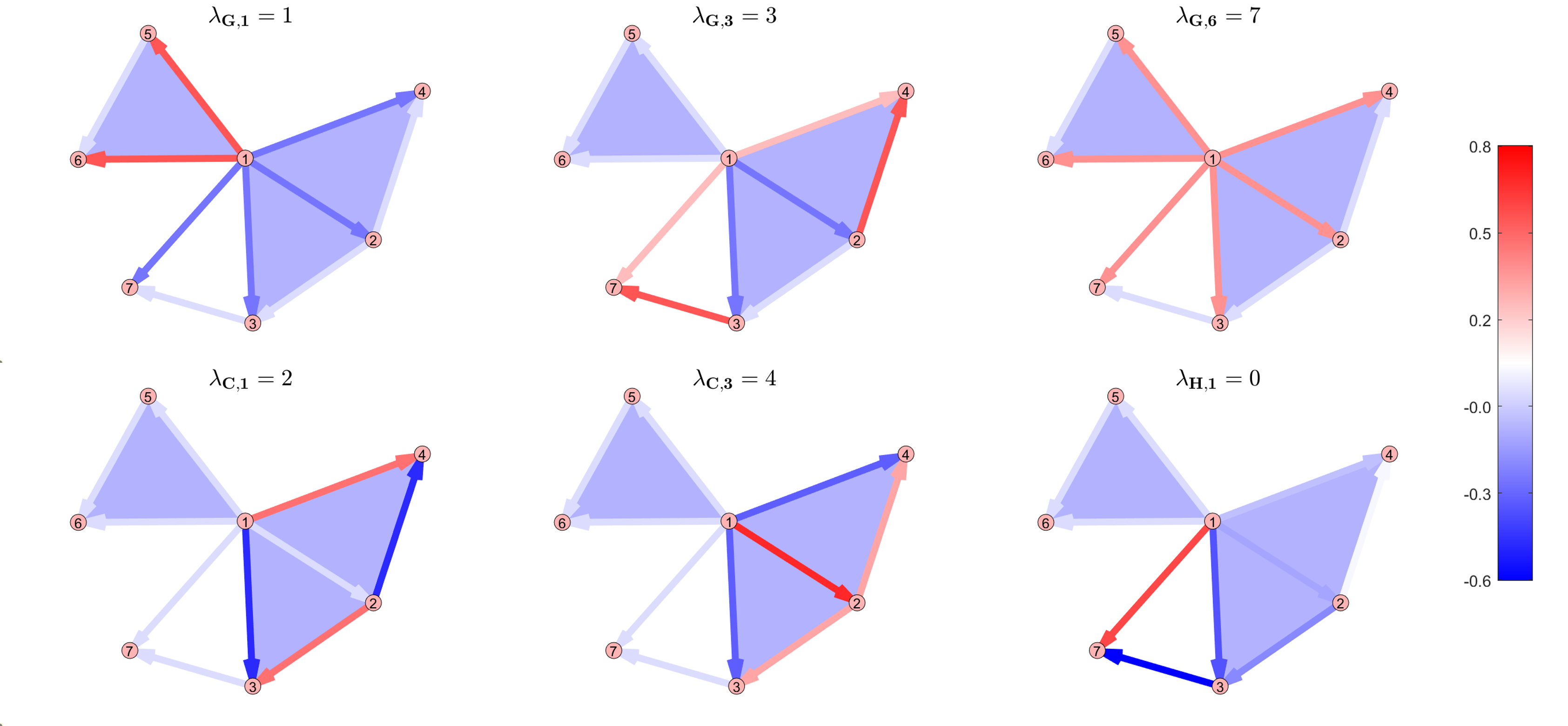}
    \caption{Eigenvectors of the $\mtL_1$ Hodge Laplacian shown as edge flow signal along with their respective eigenvalues. (top-row) gradient eigenvectors with increasing total divergence. (bottom row) curl eigenvectors with increasing total curl variation, and the harmonic eigenvector. The latter is localized around the hole in the simplicial complex.
    }
    \label{fig:variation_eigenvector}
\end{figure}

The above discussion shows that the notion of low and high frequency in a SC are only meaningful within a certain type. An illustration of the latter is shown in Figure~\ref{fig:variation_eigenvector}. This behavior is a unique characteristic of topological spaces that is not conventionally seen in graph and discrete signal processing. We refer the reader to \cite{barbarossa2020topologicalsimplicial, sardellitti2024topological, schaub2021signal, yang2022simplicial, grande2024disentangling} for more details on the latter.

\begin{remark}[Hodge decomposition and GFT] Notice that the graph Laplacian can also be written as $\mtL = \mtB_1\mtB_1^\top$, which implies that the space of all graph signals, i.e., $k = 0$, enjoys also a Hodge decomposition of the form $\fdR^{N_0} = \textnormal{span}(\mtB_1) \oplus \textnormal{kernel}(\mtL)$. Since from a topological perspective, the case $k = 0$ implies that nodes are connected only via the \emph{upper} Laplacian, the space $\textnormal{span}(\mtB_1)$ is the analogous of the curl space in \eqref{eq.HodgeDec}, whereas $\textnormal{kernel}(\mtL)$ is the harmonic space. Consequently, the GFT accounts for the projection of the graph signals onto these subspaces, where the eigenvectors $\mtU$ associated with a non-zero eigenvalue span the curl space $\textnormal{span}(\mtB_1)$ and those associated with a zero eigenvalue span the harmonic space $\textnormal{kernel}(\mtL)$. \qed
\end{remark}

\subsection{Joint Spectral Spaces}
\label{subse:dirac}


As for the Hodge Laplacians, the Dirac operator enjoys the eigendecomposition:
\begin{equation}
    \mtD_\stX = \mtD_\stX^{(d)} +\mtD_\stX^{(u)} = \mtU_\stX\mtLambda_\stX\mtU_\stX^\top.
\end{equation}
The matrix of eigenvectors $\mtU_\stX$ of $\mtD_\stX$ can be written as:
\begin{equation}\label{eq.eigendecomp}
    \mtU_\stX =  \big[\mtU_\stX^{(d)}, \mtU_\stX^{(u)}, \mttU_\stX      \big]
\end{equation}
where $\mtU_\stX^{(d)}, \mtU_\stX^{(u)}$ are the matrices of non-zero eigenvectors of $\mtD_\stX^{(d)}$ and $\mtD_\stX^{(u)}$, respectively, whose columns respectively span the joint gradient space $\textnormal{span}(\mtD_\stX^{(d)})$  and the joint curl space $\textnormal{span}(\mtD_\stX^{(u)})$. The columns of $\mttU_\stX$ span the joint harmonic space $\textnormal{kernel}(\mtD_\stX)$. Since the Dirac operator matrix is a block matrix composed of the incidence matrices defining the topological complex, it is possible to relate the eigendecomposition in \eqref{eq.eigendecomp} with the singular value decompositions of the incidence matrices $\mtB_k$,  as well as the harmonic eigenvectors of the
Hodge-Laplacian. 
Likewise, also the eigenvalues in $\mtLambda_\stX$ can be grouped into those related to $\mtU_\stX^{(d)}, \mtU_\stX^{(u)}$, and $\mttU_\stX$ and they can in turn be related to the eigenvalues of the Hodge Laplacian and singular values of the incidence matrices. Specifically, eigenvectors associated with positive eigenvalues of $\mtD_\stX^{(u)}$ and $\mtD_\stX^{(d)}$ capture configurations where
signals defined on the $1$ and $0$-simplices, and on the $2$ and $1$-simplices, respectively, are aligned with the action of the boundary operators. On the other hand, eigenvectors associated with negative eigenvalues, capture configurations where signals defined on the 1 and 0-simplices, and on the 2 and 1-simplices, respectively, are antialigned with the action of the boundary operators. By picking out the eigenvectors associated to positive (negative) eigenvalues of the Dirac operator, one can thus find appropriate basis vectors for aligned (anti-aligned) simplicial signals. The above theoretical framework and interpretation enables and motivates the design of filters acting on joint spectral spaces. We refer the reader to \cite{calmon2023dirac} for further details on the matter.

As for the topological Fourier transform, we can compute the spectral projections of simplicial signals onto the Dirac operator eigenspace. Specifically, letting $\mathbf{x}=[\mathbf{x}_0^\top,\mathbf{x}_1^\top,\mathbf{x}_2^\top]^\top\in\mathbb{R}^{N+E+T}$ be the vector collecting node, edge, and triangle signals, the topological Dirac Fourier transform reads as $\widehat{\mathbf{x}}=\mtU_\stX^\top \mathbf{x}$, where the $i$-th entry represents the weight of the $i$-th eigenvector of $\mtU_\stX$ in expressing signal $\mathbf{x}$. The inverse topological Dirac Fourier transform readily writes as $\mathbf{x}=\mtU_\stX\widehat{\mathbf{x}}$.


\section{Current Advances}
\label{sec:advances}

The goal of this section is twofold. First, it aims to bring together the current methodological advances on topological signal processing and learning. Second, it highlights how these advances tie to the concepts discussed in Sections~\ref{sec:topological} and~\ref{sec:hodge}. More specifically, in Section~\ref{subsec:spectral} we discuss topological signal representation advances, whereas in Section~\ref{subsec:learntopologies} we focus on how to learn topological structures from data. Then, in Section~\ref{subsec:filters} we discuss topological filters and in Section~\ref{subsec:sampling-reconstruction} sampling and reconstruction strategies that take into account the topological complex structure. Section~\ref{subsec:spatiotemporal_proc} discusses the case of processing spatiotemporal topological signals, and finally Section~\ref{subsec:tml} discusses how these techniques can be useful when building machine learning models for topological structures and signals.

\subsection{Signal Representation}
\label{subsec:spectral}


A spectral theory for signals defined over simplicial complexes was first introduced in \cite{barbarossa2020topologicalsimplicial}, utilizing algebraic topology tools such as higher-order Laplacian matrices. The framework identifies eigenvectors of these Laplacians as a natural basis for representing and analyzing these signals, and enables the Hodge decomposition of signals into meaningful components via spectral analysis. This decomposition identifies signals components like harmonic, gradient, and curl flows, which can then be manipulated or analyzed separately. Other studies extended the approach from \cite{barbarossa2020topologicalsimplicial} to regular cell complexes \cite{sardellitti2021topological,roddenberry2022signal}, whereas the work in \cite{sardellitti2024topological} generalized the concept of cell complexes to include hollow cells. Thease approaches to filter data on simplicial complexes operate independently on each order. To formally incorporate inter-dependencies among simplices of different orders, the Dirac operator was introduced in \cite{bianconi2021topological} and has been employed as a robust and theoretically grounded approach to jointly process signals across consecutive orders of simplices \cite{calmon2023dirac}.


A key challenge in signal processing is \textit{sparse signal representation} \cite{rioul1991wavelets}, which focuses on creating overcomplete dictionaries of atoms to represent signals as linear combinations of only a few atoms from the dictionary. Two main approaches have proven effective for both Euclidean and graph signals. The first involves analytical dictionaries, which are structured and derived from mathematical models, designed based on the specific domain while assuming a certain class of signals, such as Fourier transforms, wavelets, or curvelets. The second approach is learnable dictionaries, which are unstructured and learned from a set of training signals \cite{tovsic2011dictionary}. A crucial trade-off exists between these methods: analytical dictionaries are typically faster to implement but are less resilient to model mismatches, whereas learnable dictionaries are more robust across different signal classes but involve higher complexity due to the training phase. In the context of topological dictionaries, analytical methods have been extended from graph signals, where the topological Fourier modes provide a natural basis for signal representation \cite{schaub2020random}. However, since Fourier modes are often non-sparse and inefficient for representing localized signals, other methods have been developed. In \cite{tsitsvero2015degrees}, 
a dictionary of basis vector maximally concentrated in the discrete vertex and frequency sets were introduced, generalizing the seminal work of Slepian from continuous time to graphs. 
The work in \cite{roddenberry2022hodgelets} proposed a family of wavelets for simplicial signals that respect the Hodge decomposition. Then, the authors of \cite{battiloro2023topologicalslepians} introduced topological Slepians, which are signals maximally concentrated on the topological domain and perfectly localized in the spectral domain. Specifically, let us introduce two localization operators acting onto an edge concentration set, say $\mathcal{S}$, and onto a frequency concentration set, say $\mathcal{F}$, respectively. The edge-limiting operator onto the edge set $\mathcal{S}$ is defined as the matrix $\mathbf{C}_{\mathcal{S}}\in\mathbb{R}^{E\times E}$ given by:
\begin{align}\label{edge_limiting_operator}
\mathbf{C}_{\mathcal{S}}={\rm diag}(\mathbf{1}_{\mathcal{S}}),
\end{align}
where $\mathbf{1}_{\mathcal{S}}\in\mathbb{R}^ {E}$ is a vector having ones in the index positions specified in $\mathcal{S}$, and zero otherwise. Clearly, from (\ref{edge_limiting_operator}), an edge signal $\mathbf{x}_1$ is perfectly localized onto the set $\mathcal{S}$ if $\mathbf{C}_{\mathcal{S}} \mathbf{x}_1=\mathbf{x}_1$. Similarly, the frequency limiting operator can be defined as:   
\begin{align}\label{frequency_limiting_operator}
\mathbf{F}_{\mathcal{F}}=\mathbf{U}_1\,{\rm diag}(\mathbf{1}_{\mathcal{F}})\,\mathbf{U}^\top_1,
\end{align}
which represents an ideal band-pass filter over the frequency set $\mathcal{F}$. Clearly, an edge signal is perfectly localized over the bandwidth $\mathcal{F}$ if $\mathbf{F}_{\mathcal{F}} \mathbf{x}_1=\mathbf{x}_1$. The topological Slepians are the set of orthonormal vectors that are maximally concentrated over the edge set $\mathcal{S}$, and perfectly localized onto the bandwidth $\mathcal{F}$. Mathematically, topological Slepians are given by the solution of the following optimization problem:
\begin{align}
    \boldsymbol{\psi}_i = &\argmax_{\boldsymbol{\psi}_i}\; ||\mathbf{C}_\mathcal{S} \boldsymbol{\psi}_i||_2^2 \nonumber \\
    &\textrm{ subject to} \;\; \label{slep_prob}   ||\boldsymbol{\psi}_i|| = 1, \quad \mathbf{F}_\mathcal{F} \boldsymbol{\psi}_i = \boldsymbol{\psi}_i, \\
    & \; \qquad \qquad <\boldsymbol{\psi}_i,\boldsymbol{\psi}_j> = 0, \quad \hbox{$j = 1,\ldots,i-1$, if  $i > 1$,} \nonumber
\end{align}
for all $i=1,\ldots,E$. The solution of problem (\ref{slep_prob}) is given by the eigenvectors of the matrix operator $\mathbf{F}_\mathcal{F}\mathbf{C}_\mathcal{F}\mathbf{F}_\mathcal{F}$. Topological Slepians can be used to build localized dictionaries for signals defined over simplicial or cell complexes, which theoretically provide non-degenerate frames \cite{battiloro2023topologicalslepians}. Other generalizations are illustrated in the work \cite{saito2023multiscale}, which exploits multiscale basis dictionaries on simplicial complexes based on generalized Haar-Walsh and hierarchical graph Laplacian eigen transforms, to design multiscale hodge scattering neural networks. 

On the other side, the paper in \cite{battiloro2023topological} introduced a dictionary learning algorithm for sparse representation of signals on regular cell complexes. Utilizing Hodge theory, the dictionary is structured as polynomials of Hodge Laplacians. Specifically, the dictionary consists of the concatenation of $P$ convolutional filters, i.e.,:
$$\mathbf{H} =\{\mtH_1(\mtL_k), \mtH_2(\mtL_k), \ldots, \mtH_P(\mtL_k)\} \in \mathbb{R}^{N_k \times PN_k},$$
where each sub-dictionary is expressed as: 
\begin{equation}
\label{eq:dict_param}
    \mtH_i(\mtL_k) := \sum_{t=0}^{T_d}h_{it}^{(d)}(\mtL_{k}^{(d)})^t + \sum_{t=0}^{T_u}h_{it}^{(u)}(\mtL_{k}^{(u)})^t
\end{equation}
where $\{h_{it}^{(d)}\}, \{h_{it}^{(u)}\}$ are the filter coefficients as we discuss more in detail in Section~\ref{subsec:filters}. The dictionary learning problem, combining topological filter coefficients and sparse representation, is then efficiently solved via an alternating optimization approach.

\subsection{Learning Topologies from Data}
\label{subsec:learntopologies}

This section deals with discovering topological relationships from data. Within the field of graph signal processing, many results exist on edge-wise topology identification from node signals. Here, a link between the data and the graph is established based on concepts like smoothness, structural equation models, Gaussian graphical models, and spectral templates to name a few \cite{mateos2019connecting}. Discovering higher-order topological relationships from nodal (and potentially higher-order) observations is less obvious.

Focusing on simplicial and cell complexes, the problem statement is finding a complex that can explain the topologically data that is (fully or partially) available on different levels. In \cite{barbarossa2020topologicalsimplicial}, a hierarchical method has been proposed to infer the structure of one layer, assuming knowledge of the lower layers. For instance, if flow data (edge signals) are available, which implicitly assumes that a graph is known  (and thus ${\bf L}_0$ or ${\bf B}_1$ are known), we can infer the structure of a simplicial or cell complex of order two. Since ${\bf B}_1$ is already known, we know the lower Laplacian ${\bf L}_1^{(d)}$ and only the upper Laplacian ${\bf L}_1^{(u)}$ or ${\bf B}_2$ needs to be estimated. Referring back to the Hodge decomposition, it is clear that ${\bf B}_2$ is only needed to explain the data if the edge signals have a curl component. So first the flow data is projected onto the orthogonal complement of the gradient space. And if there is enough energy left, a fixed number of triangles is filled in such a way that some total variation measure (or smoothness measure) of the (gradient-free) flow data over the upper Laplacian ${\bf L}_1^{(u)}$ is minimized. This measure is similar to the well-known smoothness measure of node signals over the Laplacian ${\bf L}_0$. Extensions to cell complexes have been proposed in \cite{sardellitti2024topological}. In \cite{hoppe2024representing}, on the other hand, triangles are filled based on maximizing the fit of the (gradient-free) flow data to the curl space, which is different from the smoothness measure considered in \cite{barbarossa2020topologicalsimplicial,sardellitti2024topological}. 
In \cite{gurugubelli2024simplicial} and \cite{sardellitti2023probabilistic}, the observations over the elements of a simplicial complex are modeled as random variables and methods to infer the structure of the complex from data are proposed. The focus of \cite{gurugubelli2024simplicial} and \cite{sardellitti2023probabilistic} is on the edge flows, but the approach can be generalized to higher order structures. The model in  \cite{gurugubelli2024simplicial} assumes smooth Gaussian node and triangle random variables plus a white Gaussian harmonic random variable. Since the observation of edge variables implicitly assumes that the graph skeleton is known (and thus ${\bf L}_0$ or ${\bf B}_1$ is known), the problem that maximizes the posterior probability then basically boils down to a smoothness-regularized edge flow fitting problem, which can be contrasted to the maximum likelihood style fitting problem of \cite{hoppe2024representing} where no smoothness priors are assumed.
In \cite{sardellitti2023probabilistic} it was shown that, if the gradient, curl and harmonic random variables are uncorrelated, but with arbitrary covariance matrices, a principal component analysis (PCA) performed on the covariance matrix of the edge variables still satisfies the Hodge decomposition. Furthermore, simple algorithms have been proposed to infer the structure of $\mathbf{B}_2$, thus filling the triangles in the simplicial complex, based on the covariance matrix of the observed random vector projected onto the space orthogonal to the gradient space of the edge flow. A generalized version of the graphical lasso method \cite{friedman2008sparse} to random edge variables was also proposed, to estimate the partial correlation coefficients between edge variables and then infer conditional independencies among the edge variables.

Non-hierarchical methods have been proposed in~\cite{buciulea2024learninga,buciulea2024learning,wang2022fullh}. The work~\cite{buciulea2024learninga} is probably closest to the previously cited papers and jointly estimates the edges and triangles, i.e., ${\bf B}_1$ and ${\bf B}_2$, using full nodal data and partial flow data (an extension to partial nodal data is straightforward). The properties that are exploited in~\cite{buciulea2024learninga} are that the node signals are smooth over the Laplacian ${\bf L}_0$ and the edge signals have a small curl (i.e., they are smooth over the upper Laplacian ${\bf L}_1^{(u)}$). Similarly, it can be considered that the edge signals have a small divergence (i.e., they are smooth over the lower Laplacian ${\bf L}_1^{(u)}$). Or even a weighted combination of the two edge smoothness measures can be used in the objective function. On the other hand,~\cite{buciulea2024learning} relies on the Volterra model to estimate a second-order hypergraph using only nodal information. A least squares cost can be used to fit the nodal observations to the first- and second-order Volterra kernels. Furthermore, the constraint required to force the hypergraph to be a simplicial complex, i.e., to satisfy the inclusion property, can be relaxed to a convex constraint. A similar yet probabilistic Volterra-like model has been considered in \cite{wang2022fullh} for modeling social contagion dynamics. It can be viewed as an extension of the Ising model from pairwise to higher-order interactions. From observed binary infection data, the probabilities that a node or a pair of nodes is connected to another node is estimated using the expectation-maximization algorithm. Those probabilitites can then be used to recover edges and triangles of a simplicial complex.




Finally, the paper in \cite{battiloro2023latent} presents latent higher-order topology inference.
The work introduces the differentiable cell complex module (DCM), an architecture that dynamically learns cell complexes by first inferring the 1-skeleton (a graph) and then identifying higher-order cells (polygons) to include. These steps leverage message passing and a sparse sampling technique. 
The DCM is trained end-to-end and achieves significant accuracy improvements on both homophilic and heterophilic datasets, demonstrating robust performance even when input graphs poorly represent the data.


\subsection{Filtering}
\label{subsec:filters}

In Section~\ref{sec:hodge}, we discussed the spectral processing of topological signals by means of either the Hodge or Dirac decomposition. This duality between the spectrum and the topological domain opens the doors to a different signal representation and processing perspectives as discussed in Section~\ref{subsec:spectral}. Alternatively, we can spectrally process topological signals via local filtering approaches without resorting to the spectral domain. The latter is achieved in two ways: first, by extending the concept of convolutional filtering to the topological domain; and second via regularization techniques.

\smallskip
\noindent\textbf{Convolutional filtering.} The spectral filtering of topological signal is computationally expensive as we need to compute the eigendecomposition of the Hodge or Dirac matrices. To avoid the latter, \cite{yang2022simplicial} developed topological filters by following the principle of convolution to process the $k$th signal $\vcx_k$ and produce the output $\vcy_k$ as: 
\begin{equation}\label{eq.convFilt}
    \vcy_k := \mtH(\mtL_k)\vcx_k := \underbrace{\sum_{t=0}^{T_d}h_{kt}^{(d)}(\mtL_{k}^{(d)})^t}_{\mtH_k(\mtL_k^{(d)})}\vcx_k + \underbrace{\sum_{t=0}^{T_u}h_{kt}^{(u)}(\mtL_{k}^{(u)})^t}_{\mtH_k(\mtL_k^{(u)})}\vcx_k
\end{equation}
where $h_{kt}^{(d)}$ and $h_{kt}^{(u)}$ are the filter parameters associated to the lower and upper Hodge Laplacians, respectively. Here, $(\mtL_{k}^{(\cdot)})^t$ is meant as the $t$th power of the $k$th of the respective Laplacian matrix which are computed up to orders $T_d$ or $T_u$. Since the Hodge Laplacian matrices are sparse, operations of the form $(\mtL_{k}^{(\cdot)})^t\vcx_k$ can be computed recursively as $(\mtL_{k}^{(\cdot)})^t\vcx_k = \mtL_{k}^{(\cdot)}\big((\mtL_{k}^{(\cdot)})^{t-1}\vcx_k)$ to reduce the computation cost. As for the convolutional filters in time, images and graphs \cite{isufi2024graph}, the operation in \eqref{eq.convFilt} builds upon the shift-and-sum principle by propagating signal $\vcx_k$ to neighboring simplices via either the upper or lower adjacencies. This locality of information processing is the reason why the dictionaries in \eqref{eq:dict_param} are topological local dictionaries. 

The nice property of the operation in \eqref{eq.convFilt} is that is preserves a spectral analogy. In fact the lower Laplacian filtering component $\mtH_k(\mtL_k^{(d)})$ processes the gradient component of the topological Fourier transform, whereas the upper Laplacian filtering component $\mtH_k(\mtL_k^{(u)})$ processes the curl component of the TFT. This processing is coupled with that of the harmonic part via the zero-th power of the Laplacian matrices. To provide an independent processing of the Harmonic part, the work in \cite{battiloro2024generalized} modified \eqref{eq.convFilt} to:
\begin{equation}\label{eq.convFilt_harm}
    \vcy_k := \mtH(\mtL_k)\vcx_k := \sum_{t=1}^{T_d}h_{kt}^{(d)}(\mtL_{k}^{(d)})^t\vcx_k + \sum_{t=1}^{T_u}h_{kt}^{(u)}(\mtL_{k}^{(u)})^t\vcx_k + (\mtI - \epsilon\mtL_k)^{T_h}\vcx_k
\end{equation}
where the polynomials start from $t = 1$ and the term $(\mtI - \epsilon\mtL_k)^{T_h}$ projects the input $\vcx_k$ onto the harmonic space as the integer $T_h \to \infty$, if we set parameter $0 < \epsilon < 2/\lambda_{\max}(\mtL_k)$. 

When signals across multiple topological levels are present, we can readily extend operations \eqref{eq.convFilt}-\eqref{eq.convFilt_harm} onto filterbanks that account for the adjacent simplicial signals \cite{isufi2022convolutional}. To illustrate the latter, let us focus on the edge flow signals $\vcx_1$ and consider available both node signal $\vcx_0$ and triangle signals $\vcx_2$. We can then compute the output edge flow as:
\begin{equation}\label{eq.filterbank_edge}
    \vcy_1 = \underbrace{\mtH(\mtL_1)\vcx_1}_{\textnormal{edge $\to$ edge}} + \underbrace{\mtH(\mtL_1^{(d)})\mtB_1^\top\vcx_0}_{\textnormal{node $\to$ edge}} + \underbrace{\mtH(\mtL_1^{(u)})\mtB_2\vcx_2}_{\textnormal{triangle $\to$ edge}}
\end{equation}
where the node signal $\vcx_0$ and triangle signal $\vcx_2$ are transformed via the respective incidence relations onto edge flows and then processed via respective filters. Notice that the operation in \eqref{eq.filterbank_edge} can be easily extended to processing also node and triangle signals by accounting for the edge flows in a similar manner. In all these cases, the filters have different parameters so that to allow a higher flexibility when processing the different spectral components. In case the parameters are learned from limited data, we can rely on the Dirac operator \eqref{eq.Dirac} and process jointly the topological signal $\vcx_\stX = [\vcx_0, \vcx_1, \vcx_2]$ [cf. \eqref{sc_signal}] as:
\begin{equation}\label{eq.convDirac}
    \vcy_\stX = \mtH(\mtD_\stX)\vcx_\stX.
\end{equation}
Here, $\mtH(\mtD_\stX)$ is a matrix polynomial in the form \eqref{eq.convFilt} or \eqref{eq.convFilt_harm} where the  $k$th Hodge Laplacian matrix $\mtL_k$ is substituted by the Dirac matrix $\mtD_\stX$. In this case, the same scalar coefficients are used to process all simplicial signals, which allows a spectral analysis of the filter frequency response via the Dirac decomposition [cf. Section~\ref{subse:dirac}]; refer to \cite[Appendix A]{battiloro2024generalized} for more details on the latter.

\smallskip
\noindent\textbf{Filtering by regularization.} An alternative way to process topological signals with a spectral equivalence is to rely on regularization techniques in a form akin to the popular Tikhonov regularization or trend filtering for graph signals \cite{isufi2024graph}. To be more specific, let us consider the task of edge flow signal $\vcx_1$ reconstruction from partial noisy measurements $\vcf_1 = \mtM(\vcx_1 + \vcn)$, where $\mtM \in \{0, 1\}^{M_1\times N_1}$ is a masking matrix sampling $M_1 \ll N_1$ edge flows and $\vcn_1$ is a zero-mean Gaussian noise. Then, we can estimate the edge flow signal by solving:
\begin{equation}\label{eq.opt_problem}
\underset{\vcx_1 \in \mathbb{R}^{N1}}{\textnormal{argmin}}~~\|  \mtM(\vcf_1 - \vcx_1) \|_2^2 + \alpha_p \|\mtB_1\vcx_1\|_p^p + \beta_q \|\mtB_2^\top\vcx_1\|_q^q
\end{equation}
where $\alpha_p, \beta_p > 0$ are scalars and $p,q \in \{1, 2\}$. The prior information about the edge flow here is put on the regularization terms, where $\|\mtB_1\vcx_1\|_p^p$ penalizes the divergence of the edge flow to be either of low magnitude ($p = 2$) or sparse ($p = 1$), whereas the second term $\|\mtB_2^\top\vcx_1\|_q^q$ penalizes the curl component of the edge flows. Variants of problem \eqref{eq.opt_problem} have been discussed in \cite{jia2019graph,yang2022simplicialtrend,liu2023unrolling}. As for the convolutional filtering, the regularizer problem can be extended to account for multiple signal levels via the Dirac operator by penalizing either smooth or sparse differences across all the simplicial complex \cite{calmon2023dirac}.

\subsection{Sampling and Reconstruction}
\label{subsec:sampling-reconstruction}

In~\cite{barbarossa2020topologicalsimplicial}, the sampling and reconstruction problem of graph signals is extended to simplicial complexes. More specifically, an edge signal ${\bf x}_1$ is subsampled over the edge space and is further assumed to be bandlimited in the spectrum of ${\bf L}_1$, i.e., it only consists of a limited number of eigenvectors of ${\bf L}_1$. It is then shown that perfect reconstruction of ${\bf x}_1$ can be obtained under a full rank condition of the product of the edge limiting and edge-frequency limiting operators. In~\cite{barbarossa2020topologicalsimplicial}, this single-layer sampling method is further extended to multi-layer sampling, where {\em additionally} a limited number of samples of the node signal $\tilde{\bf x}_0$ and/or the triangle signal $\tilde{\bf x}_2$ forming the edge signal ${\bf x}_1$ (see~\eqref{eq.Hodgesignal}) are available, and where furthermore $\tilde{\bf x}_0$ and $\tilde{\bf x}_2$ are assumed to be bandlimited in the spectrum of ${\bf L}_0$ and ${\bf L}_2$, respectively.

In \cite{reddy2024recovery}, on the other hand, not only ${\bf x}_1$ is subsampled but also its higher-order diffusions (aggregations) over the edge Laplacian ${\bf L}_1$. The goal is then again to reconstruct this edge signal ${\bf x}_1$ but now by estimating the simplicial signals of one order lower $\tilde{\bf x}_0$ (a node signal) and one order higher $\tilde{\bf x}_2$ (a triangle signal) as well as the harmonic edge signal $\tilde{\bf x}_1$ that all three together yield the complete edge signal ${\bf x}_1$ through the Hodge decomposition (see again~\eqref{eq.Hodgesignal}). In order to do that, it is assumed that $\tilde{\bf x}_0$ and $\tilde{\bf x}_2$ are bandlimited in the spectrum of ${\bf L}_0$ and ${\bf L}_2$, respectively, and that also $\tilde{\bf x}_1$ is bandlimited over $\text{kernel}({\bf L}_1)$. As proven though in \cite{reddy2024recovery}, these bandlimitness conditions are equivalent to assuming bandlimitedness of ${\bf x}_1$ in the spectrum of ${\bf L}_1$. Hence, \cite{reddy2024recovery} is basically similar to the single-layer sampling method of~\cite{barbarossa2020topologicalsimplicial} yet replacing edge sampling by edge aggregation sampling.

Results for a product cell structure which can be viewed as the Cartesian product of two simplicial complexes can be found in \cite{reddy2024sampling}. In that context, bandlimited edge signals on the product cell complex can be written as a direct sum of the Kronecker product of bandlimited edge and node signals on the factor simplicial complexes. 

Guassian processes that encapsulate the structure of a simplicial complex have been studied in \cite{gurugubelli2023gaussian}. Edge flows are then modeled as a function of edge features, where this function has a Gaussian process distribution with a kernel that depends on ${\bf L}_1$. Based on a limited number of observed edge flows and all edge features, the kernel parameters can be learned and the edge flows on the unobserved edges can be predicted. Furthermore, an online setting can be considered where the most informative edge flows (the ones with the largest uncertainty score) are sequentially sampled.

\subsection{Bridging with Topological Machine Learning}
\label{subsec:tml}

Upon defining the notions of topological  support and local proximity between topological signals [cf. Section~\ref{sec:topological}], as well as their spectral equivalence [cf. Section~\ref{sec:hodge}], it becomes fairly natural to leverage the latter as relational inductive biases \cite{battaglia2018relational} for developing machine learning solutions for topological signals. While different reviews \cite{hajij2206topological, papillon2023architectures} and position papers \cite{papamarkou2024position} are tailored to topological deep learning, we here review works that bridge the topological signal processing techniques with the broader domain of topological machine learning.

\smallskip
\noindent\textbf{Topological neural networks (TNNs).} TNNs architectures have been proposed to learn end-to-end representations for topological signals and have been inspired by three main principles: i) message passing; ii) convolution; and iii) attention. Message passing architecture for simplicial and cell complexes have been developed in \cite{bodnar2021weisfeiler, bodnar2021weisfeilerCell}. In these cases, messages within and across different simplices or cells are exchanged and update by using the incidence matrices \eqref{inc_coeff} and the Hodge Laplacian. A principled analysis of the latter for edge flows has been proposed in \cite{roddenberry_icml_21} where the data invariance (such as permutation and orientation equivariance) are investigated to characterize the role of the simplicial proximities in each layer. A scalable solution inspired by the graph inception idea \cite{rossi2020sign} is presented in \cite{gurugubelli2024sann}. 

Convolution-based architectures have a layered structure where in each layer a convolutional filter of the forms \eqref{eq.convFilt}-\eqref{eq.convDirac} nested into pointwise nonlinearity in the same spirit as the Euclidean and graph convolutional neural networks \cite{gama2020graphs}. For example, the works in \cite{ebli2020simplicial, yang2022simplicialnn} proposed simplicial convolutional neural networks that operate at a single simplicial level, whereas extensions to signals present at different topological levels have been discussed in \cite{yang2023convolutional,zhou2024facilitating,huang2024higher}. In these cases, the simplicial convolutional filters is used to learn both the embedding from multi-hop neighboring signal information in each layer and to link the latter with the Fourier spectral interpretation discussed in Section~\ref{sec:hodge}. A bridge between convolutional and message passing solutions can be established by limiting the convolutional filter order to one, which has been discussed in \cite{chen2022bscnets, yang2022efficient} in a form akin to the popular graph convolutional networks \cite{Kipf2017-GCN}. To improve the computational efficiency of these solutions, the work in \cite{yan2025binarized} proposed a binarized simplicial convolutional neural network. Finally, the work in \cite{battiloro2024tangent} introduced a convolution operation over the tangent bundle of Riemannian manifolds, defined via the exponentials of the Connection Laplacian operator, and the corresponding Tangent Bundle Neural Networks that operate on tangent bundle signals. The authors further propose a discretization procedure for Tangent Bundle Neural Network, demonstrating that their discrete implementation constitutes a principled variant of sheaf neural networks.

Self-attention mechanisms for simplicial neural networks were concurrently introduced in \cite{giusti2022simplicial, goh2022simplicial}, with the aim of processing data components at different layers (e.g., nodes, edges, triangles, and so on), while learning how to weight both upper and lower neighborhoods of the given topological domain in a task-oriented fashion. This approach was then extended in \cite{hajij2022higher} to handle a generalized higher-order domain called a combinatorial complex. Additionally, a simplicial-based attention mechanism tailored for heterogeneous graphs was presented in \cite{lee2022sgat}. Finally, \cite{ battiloro2024generalized} introduces an attention-based neural architecture
%
for processing data on simplicial complexes using masked self-attention layers, leveraging the simplicial Dirac operator to combine data from neighboring simplices of various orders (nodes, edges, triangles) in a task-specific manner, while ensuring permutation equivariance and simplicial-awareness. Interestingly, even if typically learned attentional shift operators are not Hodge Laplacians (i.e., they do not respect the Hodge decomposition), the work in \cite{ battiloro2024generalized} proposes an attentional variant enforcing a Hodge Laplacian structure over the learned shift operators.



A way to reduce the computational complexity while improving the learning performance of TNNs is to interleave the learning layers with pooling strategies. However, to be effective and respect the topological structure of the data, these pooling layers should account for the simplicial and cell structures when coarsening the topological domain. 
An example is given in \cite{cinque2023pooling}, where the authors introduced a general simplicial pooling layer consisting of three key operations: i) a local aggregation step, which generates a meaningful summary of the input signals; ii) a selection step, which identifies an appropriate subset of simplices; and iii) a reduction step, which downsamples the input complex and the aggregated signals from step i) based on the simplices selected in step ii). 
By customizing steps ii) and iii), \cite{cinque2023pooling} proposed four distinct simplicial pooling layers that extend the widely used graph pooling strategies. Another interesting approach was then proposed in \cite{mcguire2023nervepool}, which introduced a pooling layer for simplicial complexes that uses vertex partitions to create hierarchical representations, collapsing information through learned vertex cluster assignments and deterministically coarsening higher-dimensional simplices.


\smallskip
\noindent\textbf{Gaussian processes (GPs).} Gaussian processes are a class of statistical models that can quantify the uncertainity associated with their prediction. In essence, GPs are distribution over functions characterized by the mean function and a covariance kernel function that models the dependence between function values at two different inputs. When these inputs are topological signals, we would like to include their topological structure into the covariance matrix and define appropriate kernels to aid learning. The work in \cite{yang2023hodge} discussed this case when said kernels have a spectral meaning and can distinguish between the gradient and curl components of the signal. A related discussion is present in \cite{alain2023gaussian} for cell complexes. Finally, topologically-aware GPs for edge flow reconstruction and simplicial closure are discussed in \cite{gurugubelli2023gaussian, gurugubelli2024gaussian}, respectively.


\smallskip
\noindent\textbf{Data augmentation and self-supervised learning.} The particular coupling between topological signals and their underlying structure can be used for data augmentation purposes; and in particular to train TNNs in a self-supervised manner. In this regard, the work in \cite{navarro2024sc} builds upon the mixup technique \cite{zhang2017mixup} to generate artificial topologies that respect the simplicial complex structure. A topological-domain augmentation perspective is also discussed in \cite{madhu2024toposrl} where random augmentations that respect the simplicial structure are used as data generation for self-supervised learning of TNNs. Differently, \cite{mollers2024hodge} discuess data augmentation techniques for SSL that respect spectral prior and uses the latter as an additional bias to aid learning. 

\subsection{Processing Spatiotemporal Signals}
\label{subsec:spatiotemporal_proc}

The methods discussed so far focus on time-invariant signals over topological spaces. However, in many real-world applications, observations typically consist of time series defined on various higher-order structures, including edges, triangles, and beyond \cite{schaub2021signal,Volterra2021} that influence each other in a non-trivial manner. As an example, in water networks, the water flow in the pipes can be better modeled as a time-varying process evolving over the edges of the network, whereas the pressure is a process over the nodes. The mutual influence between these edge and vertex signals underscores the need for learning models that can capture their dependencies across both time and the higher-order structure. To account for the latter spatiotemproal coupling, a few works have emerged that either rely on recurrent models, adaptive filtering, or product spaces.


\smallskip
\noindent\textbf{Recurrent models.} One of the key recurrent models for multivariate data is the vector autoregressive model (VAR) \cite{lutkepohl2005new,onlzam2019,RohanDSLW,RohanMLSP,RohanTSP,nonshe2019,veedu2021topology,GVARElvin}. While this model captures the temporal dependencies among the time-varying processes as a linear combination of their past realizations, it ignores the underlying network structure of the data and, consequently, do not grasp the inductive biases inherent in them. Given that the network topology affecting the process is typically sparse, a standard VAR model ignoring this structure fails to capitalize on the sparsity of interactions between the time series, suffering from the curse of dimensionality. This disadvantage can be alleviated with the introduction of network-based VAR models, such as the graph VAR (G-VAR) model \cite{GVARElvin}, which considers the time series as processes over the vertices of a graph and models their evolution as a sparse linear combination of the time series in the adjacent vertices. It accounts for the structure and captures the sparsity in the data via the so-called graph convolutional filters \cite{Grpah_Filter_Elvin}, enabling parameter sharing across vertices. This results in a computational complexity of a G-VAR model is linear in the number of time series, while that of standard approaches is quadratic \cite{onlzam2019,RohanTSP,facotor_model,Shrinkage,dim_red,extra3,extra4,extra5}. Similarly, it is also possible to learn G-VAR models over signals defined exclusively on the edges of a graph (e.g. flows of a water network). However G-VAR models do not exploit higher-order dependencies.

In \cite{Krishnan_2024}, SC-VAR models are introduced to represent time series defined over higher-order networks. Leveraging the Hodge Laplacian representation of simplicial complexes and simplicial convolution filters, the proposed model efficiently captures temporal interactions among signals across different higher-order network structures, while attaining orders of magnitude less parameters than conventional VAR models. The SC-VAR model can capture spatio-temporal dependencies between signals defined on distinct simplicial levels, as we describe next. 

To be more specific, let us consider a time varying vertex signal $\vcx_{0,t}$, edge signal $\vcx_{1,t}$, and triangle signal $\vcx_{2,t}$. By leveraging the adjacencies between simplices of different orders, we can process a $k-$simplicial signal $\bbx_{k,t-p}$ by first filtering it with a filter $\bbH_{k,p}(\bbL_k)$, then transforming it into its adjacent simplex via the corresponding incidence matrix, and finally filtering the transformed signal via another filter  $\bbG_{k,p}(\bbL_{k+1})$ or $\bbG_{k,p}(\bbL_{k-1})$, e.g., $\bbG_{k,p}(\bbL_{k+1})\bbB_k^\top\bbH_{k,p}(\bbL_k)\bbx_{k,t-p}$. Following, this \emph{convolve-transform-convolve} principle, we define a SC-VAR model of order $P$ as:
\begin{align}\label{eq.SCVAR}
\begin{split}
\bbx_{0,t} &= \sum_{p = 1}^P\bbH_{p}^{00}(\bbL_0)\bbx_{0,t-p}+ \bbG_{p}^{01}(\bbL_0)\bbB_1\bbH_{p}^{01}(\bbL_1)\bbx_{1,t-p} + \bbn_{0,t}\\
\bbx_{1,t} &= \sum_{p = 1}^P\bbG_{p}^{10}(\bbL_1)\bbB_1^\top\bbH_{p}^{10}(\bbL_0)\bbx_{0,t-p}	\!+\!\bbH_{p}^{11}(\bbL_1)\bbx_{1,t-p}\\
&\qquad\qquad  \!+\! \bbG_{p}^{12}(\bbL_1)\bbB_2\bbH_{p}^{12}(\bbL_2)\bbx_{2,t-p} \!+\! \bbn_{1,t}\\
\bbx_{2,t} &= \sum_{p = 1}^P\bbG_{p}^{21}(\bbL_2)\bbB_2^\top\bbH_{p}^{21}(\bbL_1)\bbx_{1,t-p}+ \bbH_{p}^{22}(\bbL_2)\bbx_{2,t-p} + \bbn_{2, t}
\end{split}
\end{align}
where $\bbn_{k,t}$ denotes the model error.
%
The SC-VAR expression in \eqref{eq.SCVAR} consists of three $k-$process equations, each for $k=0,1,~\text{and},2$, delineating the processes related to vertices, edges, and triangles. The indices $m$ and $n$ of filters $\bbG_{p}^{mn}(\cdot)$ and $\bbH_{p}^{mn}(\cdot)$ represent the simplicial levels after and before the convolve-transform-convolve operations, i.e., we are processing a $n-$simplicial signal on the $m-$process equation. When processing a $k-$signal on the $k-$process equation, the \emph{transform} operation is not required, and hence such terms in \eqref{eq.SCVAR} do not require the post-transform filters $\bbG_p^{kk}(\cdot)$. 

%
%
%
%
%
A particular instance of the SC-VAR model is the Simplicial-VAR (S-VAR) model, which neglects the vertical adjacencies, overlooking the dependencies across distinct simplicial levels: 
\begin{align}\label{eq.SVAR}
\bbx_{k,t} = \sum_{p = 1}^P\bbH_{k,p}(\bbL_k)\bbx_{k,t-p} + \bbn_t, ~~\text{for}~~k = 0, 1, 2,
\end{align}\\[-4mm]
where {\small$\{\bbH_p^k(\bbL_k)\}_{p=1}^P$} are simplicial convolutional filters \cite{Joshin_SVAR}. 

An online learning method is proposed in order to update the time-varying SC-VAR model parameters of the model from streaming time series, providing convergence guarantees in terms of an upper bound for the dynamic regret, attaining a sublinear dynamic regret under reasonable assumptions suitable for real-world applications. In addition, a joint simplicial-temporal Fourier transform is also introduced by extending the simplicial Fourier transform concept  \cite{barbarossa2020topological} and analyzing the SC-VAR model in the spectral domain to investigate the learned frequency responses of the model. Unlike the analogous G-VAR analysis \cite{GVARElvin}, a coupled two-dimensional frequency response is shown between different types of simplicial frequencies.

Both SC-VAR and S-VAR have been representations that have been exploited in several relevant problems, such as online edge flow imputation, evolution backcasting of edge flows, spatio-temporal filtering and Kalman filtering for simplicial processes. An algorithm to retroactively compute the evolution of edge signals is proposed in \cite{Rohan_backcasting2024}. In this case, first, this method learns an S-VAR representation and estimates missing edge signals from partially observed data using a block coordinate descent technique; and then, leveraging the learned S-VAR representation and the reconstructed signals, the algorithm backcasts the evolution of edge flows prior to the partial observations. It should be noticed that backcasting becomes more challenging when we only possess partial observations, which is common in practical scenarios (e.g. traffic or water networks) where data availability is limited due to sensor and communication failures, or due to the impracticality of placing sensors in some locations. However, the rich structural information rooted in the simplicial complex structure, enables the algorithm to address this effectively.

In \cite{nguyen2024}, the simplicial-complex structure and Hodge Laplacians are exploited to incorporate the inductive bias in combination with multilinear kernel regression to perform edge imputation via manifold learning (MultiL-KRIM) \cite{NguyenOJSP2024}. MultiL-KRIM assumes that missing entries can be estimated by a set of landmark points, extracted from measurements
and located around a smooth manifold, embedded in an ambient reproducing kernel Hilbert space (RKHS), which allows to perform functional approximation and perform imputation of time-varying edge flows. Similarly, in \cite{Rohan_SPL23}, an online algorithm is presented for time-varying edge flow imputation which combines learning a line graph identification via a vector autoregressive model and a group-Lasso-based optimization with a Kalman filtering-based reconstruction, leveraging a simplicial complex (SC) representation over the underlying learned causal dependencies.
Another Kalman filtering approach for simplicial processes has been recently presented in \cite{RohanAsilomar2024}, where it is assumed that the hidden dynamics of a system can be expressed as a simplicial process that
respects the structure of the underlying network topology. These dynamics
are observed through an observation matrix, which is represented using
simplicial convolution filters. The combination of linear dynamics
and linear observation enables the use of a Kalman filter
to compute the best linear unbiased estimate of the process.
Additionally, a parametric, structure-aware noise
covariance model is proposed for the system dynamics, and the algorithm alternates between
estimating the process state using the Kalman filter and updating
the parameters through maximum likelihood estimation. 

While these methods focus primary on linear models, the work in \cite{chen2022time} proposed a recurrent neural network type TNN perspective for modeling nodal time series, where the simplicial complexes represent the higher-order interactions of the hidden states.

%

\smallskip
\noindent\textbf{Adaptive filtering.} Topological filtering can also be cast in the context of adaptive learning, whose aim is to infer and track the structure of an unknown system from streaming and noisy data observed over time. As an example, the work in \cite{marinucci2024topological} introduced a topological least-mean squares algorithm to process and learn from streaming signals
defined over cell complexes. Specifically, let $\mathbf{x}_t$ be stationary edge flow signals\footnote{We here drop the subscript $k= 1$ to avoid overcrowded symbols.}, 
processed over time by the linear shift-invariant graph filters  according to the following model: 
\begin{align}\label{eq:lin_obs}
    \mathbf{y}_t = \mathbf{M}_t\Bigg[\sum_{m=1}^{T_d}h_{m}^{(d)}(\mtL^{(d)})^m \mathbf{x}_{t-m} 
    +\sum_{m=0}^{T_u}h_{m}^{(u)}(\mtL^{(u)})^m \mathbf{x}_{t-m} + \mathbf{n}_t\Bigg],    
\end{align}
with $t \geq \max\{T_d, T_u\}$, where $\vcn_t$ 
denotes an 
i.i.d. zero-mean measurement noise, independent of any other signal; 
$\mathbf{h}$ 
is the vector collecting all topological filter coefficients to be estimated; and $\mathbf{M}_t={\rm diag}(m_{1t},\ldots,
m_{M_1,t})
\in\mathbb{R}^{M_1\times M_1}$ is a sampling operator such that $m_{it}=1$ if edge $i$ is sampled at time $t$, and zero otherwise. The goal is to estimate the filter coefficients \(\mathbf{h}\) in (\ref{eq:lin_obs}) using a mean-square-error criterion:
\begin{align}\label{eq:MSE}
   \min_{\mathbf{h}}\;\;\mathbb{E}\{\big\|\mathbf{y}_t - \mathbf{M}_t\mathbf{X}_t\mathbf{h}\big\|^2 \},
\end{align}
where matrix 
$\mathbf{X}_t$ is given by: 
\begin{align}\label{eq:X}
    \mathbf{X}_t = \big[\mathbf{x}_t, \mathbf{L}_d \mathbf{x}_{t-1},\ldots,\mathbf{L}_d^{T_d}\mathbf{x}_{t-T_d},
     \mathbf{L}_u \mathbf{x}_{t-1},\ldots,\mathbf{L}_u^{T_u}\mathbf{x}_{t-T_u}
    \big],
\end{align}
which collects shifted versions of the edge flows $\mathbf{x}_{t-m}$ over the upper and lower neighborhoods. Then, parameters $\mathbf{h}$ can be determined by proceeding iteratively using stochastic gradient descent:
\begin{align}\label{eq:grad_desc}
   \mathbf{h}_{t+1} = \mathbf{h}_t + \mu \mathbf{X}^\top_t\mathbf{M}_t\big( \mathbf{y}_t - \mathbf{X}_t \mathbf{h}_t \big), 
\end{align}
with $\mu>0$ denoting a (sufficiently small) step-size. Recursion (\ref{eq:grad_desc}) is referred to as the topological LMS algorithm \cite{marinucci2024topological}, where streaming flow signals $\{\mathbf{y}_t,\mathbf{x}_t\}$ are processed taking into account the topological information coming from the cell complex domain.

\smallskip
\noindent\textbf{Product spaces.} An alternative to the above serial models, is to build upon the concept of product graphs \cite{sandryhaila2014big,sabbaqi2023graph} and enable spatio-temporal filtering of time-varying signals on SCs. The work in \cite{Roddenberry2022} presents a framework for signal processing on product spaces of simplicial complexes, where the time axis is interpreted as a SC, and a cellular complex is constructed to represent the spatially distributed, 
time-varying signal. It leverages the structure of the
eigenmodes of the Hodge Laplacian of the product space to jointly
filter along time and space. For this, they use a decomposition 
of the Hodge Laplacian of the product space, which shows how the 
product structure induces a decomposition of each eigenmode into 
a spatial and temporal component. This technique is applied 
to interpolate trajectories from a limited set of observed trajectories.

\section{Applications}
\label{sec:applications}

While topological signal processing and learning is an emerging research direction, they have shown promise in some key application areas or lend themselves in developments in different fields. Complementary application areas tailored to topological deep learning can be found in \cite{papamarkou2024position}. 

\smallskip
\noindent\textbf{Processing network flow signals.} Flow signals appear in a myriad of infrastructure networks such as water, power, transportation and telecommunication networks, among others. Such signals reside naturally on the edges of a network and are coupled with the node signals. One key challenge in critical infrastructure networks is that the overall network state (node+edge signals) need be estimated or forecasted by a few observations. The topological sampling techniques discussed in Section~\ref{subsec:sampling-reconstruction} can be used for sensor placement. One such case has been discussed in \cite{kerimov2024sensor} for sensor placement in water distribution systems, where topological Gaussian processes have also been exploited. Additionally, topologically-aware interpolation strategies can be used for state estimation, which consists of in a node-and-edge signal interpolation task. In the water network case, a topological neural networks have been used in \cite{kerimov2024towards} to develop surrogate models that can transfer to unseen networks. Often, in these cases we may often require inferring edge flows from partial nodal data; in this case the physics of these models need be taken into account where the Hodge theory can be used to map between node and edge flows. The latter idea was used to develop implicit layers for simplicial neural networks in \cite{smith2022physics}. Simplicial neural networks and Hodge representation of networks has also been used in \cite{chen2023learning} for power outage detection and in \cite{xia2024locational} for false data injection attacks in smart grids. When such anomalies can be localized on particular Hodge subspaces [cf. Section~\ref{sec:hodge}], a mathematically tractable matched subspace detector with optimality guarantees \cite{liu2024hodge} can be used in place of neural network solutions. A further application of edge flow processing comes from the analysis of a discrete vector field, defined as a set of vectors associated to a point loud. As shown in \cite{barbarossa2020topologicalsimplicial}, operating a Delaunay triangulation of the point cloud, a discrete vector field can be converted into a scalar field obtained by projecting the vector on each vertex onto the incident edges of the triangulation. The resulting edge flow can then be filtered using the methods described in the previous sections. The resulting scalar edge signal can then be mapped back onto a filtered discrete vector field living on the original point cloud.

\smallskip
\noindent\textbf{Geometry processing and vector calculus.} Historically, point cloud processing has been approached by graph-based techniques on the 3D mesh \cite{hu2021graph}. Such a paradigm has in fact been the early roots of developing graph signal processing techniques \cite{taubin1995signal,taubin1996optimal} till the latest progress with graph neural networks \cite{wu2020comprehensive}. More recently, a topological-based mesh processing is taking place that combine vector calculus with mesh processing via the Hodge theory. For example, vector field-based computational processing in surfaces plays a crucial role in encoding both direction and sizing of the surfaces. By means of vector calculus and Hodge-Helmholtz decomposition it is possible to characterize and process the coupling of surface data on nodes, edges and triangular faces, ultimately, linking topological signal processing and learning with differential geometry \cite{crane2013digital, de2016vector}. For example, \cite{brandt2017spectral} considers a Hodge spectral processing of tangential vector fields that has strong ties with the topological Fourier analysis in Section~\ref{sec:hodge}. This spectral-based perspective for geometry processing has also been used in \cite{keros2023spectral} to develop a spectral coarsening approach to expedite processing. Such spectral duality has also been used in end-to-end learning via tailored topological neural networks that are both geometric- and Hodge-aware in \cite{smirnov2021hodgenet,wiersma2022deltaconv}. We hypothesize that bridging advances in geometry processing with those topological signal processing and learning can expedite the method development in both domains, in the same way, mesh-based and graph-based processing complemented each other.

\smallskip
\noindent\textbf{Statistical ranking.} Hodge theory and decompositions have shown great potential in statistical ranking of lists \cite{jiang2011statistical}. In particular, each vertex is an item in a list and an edge flow is considered as a ranking order; i.e., if the flow goes from vertex $i$ to $j$ then the ranking score is higher at $i$ than $j$. Then, via the Hodge decomposition, a gradient flow shows a global consistency in ranking as they always go from higher-to-lower ranking scores, whereas a presence of an harmonic or curl flow would show local or global inconsistencies in rankings. This Hodge-based statistical ranking (HodgeRank) has been successfully applied to top-N recommender system lists and currency exchange markets \cite{jiang2011statistical} as well as in biomolecular data  analysis \cite{wei2022hodge}. The TSPL methods can further aid statistical ranking. For example, they can be used to remove ranking noise, perform ranking with missing values, or sample a minimum number of items to guarantee a certain ranking consitency. TNNs, and in particular those linked with the Hodge spectrum such as convolutional architectures \cite{yang2022simplicialnn} can be used to learn deep statistical ranking models.

\smallskip
\noindent\textbf{Biomolecular data.} A key challenge in graph-based learning is to classify molecules or even synthesize new one from a limited training set \cite{wieder2020compact,zhang2023survey}. In fact, many of the message-passing TNN advances have been developed to overcome the limited Weisfeiler-Lehman expressivity of GNNs in graph classification tasks \cite{bodnar2021weisfeiler, bodnar2021weisfeilerCell, gurugubelli2024sann}. Molecules in particular can be seen as structurally rich graphs with (hidden) topological information that are combined their features \cite{jiang2021topological}. By inducing higher-order simplicial or cell structures within a molecular graph --a technique known as lifting \cite{bernardez2024icml}-- TNNs can leverage these topological relations to learn more expressive representations. The Hodge theory discussed in Section~\ref{sec:topological} and Section~\ref{sec:hodge} has been used in \cite{wei2022hodge} to analyze biomolecular structures. More specifically, these biomolecular structures have been represented via simplicial complexes with certain edge flows, and the spectrum of the respective Hodge Laplacian reveals particular properties of the structure as well as characterizes folding or compactness of the biomolecule. Linking the latter with flow variabilities and the topological Fourier transform represents a promising unexplored direction that may reveal deeper insights on the relation between molecular features and structure.

\smallskip
\noindent\textbf{Brain networks.} Topological processing tools have been also largely applied to represent and extract information from brain network data. Specifically, the work in \cite{lee2019coidentification} addresses the challenge of identifying shared topological (group-level hole) substructures in brain networks by extending graph Laplacians to higher-order Hodge Laplacians. 
Then, the study in \cite{anand2023hodge} employs Hodge decomposition to analyze brain networks by breaking them into gradient, curl, and harmonic flow components, effectively capturing complex topological features. Using a Wasserstein distance-based topological inference, the method reveals statistically significant differences in the topological properties of male and female brain networks from resting-state fMRI data. The works in \cite{dakurah2022modelling,anand2024hodge} presents a persistent homology-based framework using the Hodge Laplacian to extract and analyze cycles in brain networks, validated through simulations and resting-state fMRI data. The authors of \cite{nasrin2019bayesian} introduced a noise-resilient method for EEG analysis using persistent homology and a Bayesian framework, enabling effective classification of noisy, nonlinear, and nonstationary signals. 
Other approaches have augmented GNNs with topological information \cite{huang2023heterogeneous,park2023convolving,hwang2024multi}. For instance, the study \cite{huang2023heterogeneous} introduces a novel heterogeneous graph convolutional neural network using Hodge-Laplacian operators and topological graph pooling to analyze fMRI data, outperforming state-of-the-art GNNs in predicting general intelligence and providing interpretable neural circuit insights. Similarly, the work \cite{park2023convolving} proposed Hodge-Graph Neural Network, which leverages the Hodge Laplacian to directly model edgewise topological features in brain networks, outperforming traditional GNNs and graph classification models in the Alzheimer’s Disease Neuroimaging Initiative study. Finally, the authors \cite{hwang2024multi} introduces a dual aggregation framework for brain network analysis that simultaneously learns node and edge embeddings, capturing intricate node-edge relationships and demonstrating superior performance and interpretability.



\smallskip
\noindent\textbf{Epidemic modeling.} Modeling and predicting epidemic spreading represents one of the cornerstone applications network science \cite{pastor2015epidemic}. This problem is conventionally seen as a dynamic process over the nodes of the graph, where the signal on the nodes indicates the state and edges represent probabilities of infection (e.g., susceptible vs. infected). Bridges of the latter with graph-based processing techniques can be found in \cite{yan2017graph,tomy2022estimating,liu2024review}. More recently, contagion and spreading processes in complex systems have been better understood via higher-order networks and simplicial complexes; see e.g., \cite{iacopini2019simplicial,battiston2020networks,battiston2021physics}. While direct bridges between TSPL and epidemic spreading have yet to be established, we hypothesise that many of the advances presented in this paper can facilitate the latter and open the doors to new insights.

\smallskip
\noindent\textbf{Semantic communication.} A key challenge in semantic communications is managing the complexity of data representation while ensuring transmitted symbols effectively convey intended meanings within acceptable distortion or perceptual variation \cite{barbarossa2023semantic,strinati2024goal}. This demands robust methods to model, extract, and encode data semantics. Although a universal definition of semantics in communication is still lacking, it can be understood as the relationships between elements of a broadly defined language, often formalized using topological spaces—mathematical structures describing elements and their relations \cite{barbarossa2023semantic}. Defining multi-way relationships allows data to be represented as signals over a topological space, where TSP becomes crucial for semantic data representation and communication. In particular, the choice of the semantic topological space directly impacts our ability to process the data and achieve an efficient, parsimonious representation that optimally balances distortion, perception, and complexity of data representation.



\section{Conclusion and Open Issues}
\label{sec:OpenIssues}

This paper brought together recent methodological advances in topological signal processing and learning. Instrumental to these methods is to represent the higher-order data structure via simplicial or cell complexes, which consists of well-structured richer forms of hypergraphs. By representing these topological structures via the Hodge Laplacians, it is possible to generalize key signal processing concepts from the Euclidean and graph domain to topologies, where now we can characterize signals defined on nodes, edges and other higher-order topological structures such as triangular faces. 
Relying on the latter and on the spectral theory of these Hodge Laplacians, we highlighted striking differences between topological Fourier domain and Euclidean and graph Fourier domain. Subsequently, methodological advances have been revisited and linked to potential application domains where graph-based techniques remain limited.

While recent works have shown their promise for some important applications, there remain significant open issues. 

\begin{itemize}
    \item First, on the topological representation front, these methods rely predominantly on undirected and unweighted topological structures. This, however, limits drastically their representation power as dependency between higher-order signals can be better represented as a directed or a weighted one. Some recent advances about weighted representation of topological complexes have been proposed in \cite{baccini2022weighted,battiloro2023topological} and the first works have emerged about algebraic representation of directed topologies \cite{riihimaki2023simplicial, gong2024higher} and directed topological neural networks \cite{lecha2024higher}. Yet, these directions remain largely unexplored and the impact of weights and directionality on TSPL methods is little studied.
    \item Second, inferring the simplcial and cell structure from data poses not only the conceptual challenge of how to tie the signal with the topology, but also poses a large computational challenge. While some techniques have emerged as discussed in Section~\ref{subsec:learntopologies} scaling the latter to hundreds of thousands of nodes as may be the case in brain structures is non-trivial.
    \item Third, the spectral techniques for TSPL are meaningful when processing signals at a particular topological level. When multiple signals are present, we can rely on spectrum of the Dirac operator [cf. Section~\ref{subsec:hodgeDec}] but the latter is not as meaningful as the spectrum of the Hodge Laplacians. Especially, for spectral analysis of topological signals deeper insights on joint spectral processing of multiple topological signals remains an open issue.
    \item Fourth, most of the datasets and applications contain signals up to edge flows, ultimately, limiting the potential of TSPL methods. Identifying, significant application areas with data over second or higher order simplicial/cell structures may be a substantial leap forward towards developing more powerful TSPL methods. Links of the latter with topological data analysis and topological deep learning can further accelerate the impact.
    \item Fifth, on the application front, TSPL methods has shown the most promise for signals up to edge flows present in data sets exhibiting irregular relationships that can captured by simplicial or cell strucrures. However, additional research is needed to identify promising application areas where signals reside naturally on higher-order simplicial or cell structures. The latter will not only open the doors to new application domains but will also provide deeper insights on the advantages and limitations of these techniques for handling irregular data.
\end{itemize}

\section*{Acknowledgements}

The work of E. Isufi and G. Leus has been partly supported by the the Dutch Research Foundation, Open Technology Programme GraSPA proposal $\#19497$. The work of E. Isufi has been also partly supported by the TU Delft AI Labs programme and the Dutch Research foundation VENI proposal $\#19052$.  
The work of B. Beferull-Lozano has been partly supported by the IKTPLUSS DISCO grant 338740, Research Council of Norway.

The work of S. Barbarossa and P. Di Lorenzo has been partly supported by European Union under the Italian National Recovery and Resilience Plan of NextGenerationEU, partnership on
Telecommunications of the Future (PE00000001 - program RESTART). The work of P. Di Lorenzo has been also partly funded the SNS JU project 6G-GOALS under the EU’s Horizon program Grant Agreement No 101139232.



\bibliographystyle{elsarticle-num} 
\bibliography{bibFiles/biblioFilters}

\begin{thebibliography}{100}
\expandafter\ifx\csname url\endcsname\relax
  \def\url#1{\texttt{#1}}\fi
\expandafter\ifx\csname urlprefix\endcsname\relax\def\urlprefix{URL }\fi
\expandafter\ifx\csname href\endcsname\relax
  \def\href#1#2{#2} \def\path#1{#1}\fi

\bibitem{karlebach2008modelling}
G.~Karlebach, R.~Shamir, Modelling and analysis of gene regulatory networks,
  Nature reviews Molecular cell biology 9~(10) (2008) 770--780.

\bibitem{jackson2008social}
M.~Jackson, Social and economic networks (2008).

\bibitem{richiardi2013machine}
J.~Richiardi, S.~Achard, H.~Bunke, D.~Van De~Ville, Machine learning with brain
  graphs: predictive modeling approaches for functional imaging in systems
  neuroscience, IEEE Signal processing magazine 30~(3) (2013) 58--70.

\bibitem{newman2018networks}
M.~Newman, Networks, Oxford university press, 2018.

\bibitem{lauritzen1996graphical}
S.~Lauritzen, Graphical models, Clarendon Press, 1996.

\bibitem{ortega2018graph}
A.~Ortega, P.~Frossard, J.~Kova{\v{c}}evi{\'{c}}, J.~M.~F. Moura,
  P.~Vandergheynst, Graph signal processing: Overview, challenges and
  applications 106~(5) (2018) 808--828.

\bibitem{bronstein2021geometric}
M.~M. Bronstein, J.~Bruna, T.~Cohen, P.~Veli{\v{c}}kovi{\'c},
  \href{http://arxiv.org/abs/2104.13478}{Geometric deep learning: Grids,
  groups, graphs, geodesics, and gauges}, arXiv:2104.13478v2 (2022).
\newline\urlprefix\url{http://arxiv.org/abs/2104.13478}

\bibitem{isufi2024graph}
E.~Isufi, F.~Gama, D.~I. Shuman, S.~Segarra, Graph filters for signal
  processing and machine learning on graphs, IEEE Transactions on Signal
  Processing (2024).

\bibitem{sandryhaila2014discrete}
A.~Sandyhaila, J.~M.~F. Moura, Discrete signal processing on graphs: Frequency
  analysis 62~(12) (2014) 3042--3054.

\bibitem{tsitsvero2016signals}
M.~Tsitsvero, S.~Barbarossa, P.~Di~Lorenzo, Signals on graphs: Uncertainty
  principle and sampling, IEEE Transactions on Signal Processing 64~(18) (2016)
  4845--4860.

\bibitem{tanaka2020sampling}
Y.~Tanaka, Y.~C. Eldar, A.~Ortega, G.~Cheung, Sampling signals on graphs: From
  theory to applications, IEEE Signal Processing Magazine 37~(6) (2020) 14--30.

\bibitem{battaglia2018relational}
P.~W. Battaglia, J.~B. Hamrick, V.~Bapst, A.~Sanchez-Gonzalez, V.~Zambaldi,
  M.~Malinowski, A.~Tacchetti, D.~Raposo, A.~Santoro, R.~Faulkner, et~al.,
  Relational inductive biases, deep learning, and graph networks, arXiv
  preprint arXiv:1806.01261 (2018).

\bibitem{wu2020comprehensive}
Z.~Wu, S.~Pan, F.~Chen, G.~Long, C.~Zhang, S.~Y. Philip, A comprehensive survey
  on graph neural networks, IEEE transactions on neural networks and learning
  systems 32~(1) (2020) 4--24.

\bibitem{grady2010discrete}
L.~J. Grady, J.~R. Polimeni, Discrete calculus: Applied analysis on graphs for
  computational science, Vol.~3, Springer, 2010.

\bibitem{lambiotte2019networks}
R.~Lambiotte, M.~Rosvall, I.~Scholtes, From networks to optimal higher-order
  models of complex systems, Nature physics 15~(4) (2019) 313--320.

\bibitem{majhi2022dynamics}
S.~Majhi, M.~Perc, D.~Ghosh, Dynamics on higher-order networks: A review,
  Journal of the Royal Society Interface 19~(188) (2022) 20220043.

\bibitem{bick2023higher}
C.~Bick, E.~Gross, H.~A. Harrington, M.~T. Schaub, What are higher-order
  networks?, SIAM Review 65~(3) (2023) 686--731.

\bibitem{berge1984hypergraphs}
C.~Berge, Hypergraphs: combinatorics of finite sets, Vol.~45, Elsevier, 1984.

\bibitem{zhang2019introducing}
S.~Zhang, Z.~Ding, S.~Cui, Introducing hypergraph signal processing:
  Theoretical foundation and practical applications, IEEE Internet of Things
  Journal 7~(1) (2019) 639--660.

\bibitem{pena2023t}
K.~Pena-Pena, D.~L. Lau, G.~R. Arce, T-hgsp: Hypergraph signal processing using
  t-product tensor decompositions, IEEE Transactions on Signal and Information
  Processing over Networks 9 (2023) 329--345.

\bibitem{zhang2020hypergraph}
S.~Zhang, S.~Cui, Z.~Ding, Hypergraph spectral analysis and processing in 3d
  point cloud, IEEE Transactions on Image Processing 30 (2020) 1193--1206.

\bibitem{pena2023learning}
K.~Pena-Pena, L.~Taipe, F.~Wang, D.~L. Lau, G.~R. Arce, Learning hypergraphs
  tensor representations from data via t-hgsp, IEEE Transactions on Signal and
  Information Processing over Networks (2023).

\bibitem{wang2024t}
F.~Wang, K.~Pena-Pena, W.~Qian, G.~R. Arce, T-hypergnns: Hypergraph neural
  networks via tensor representations, IEEE Transactions on Neural Networks and
  Learning Systems (2024).

\bibitem{lim2020hodge}
L.-H. Lim, Hodge laplacians on graphs, Siam Review (2020).

\bibitem{shuman2013emerging}
{D. I Shuman}, S.~K. Narang, P.~Frossard, A.~Ortega, P.~Vandergheynst, The
  emerging field of signal processing on graphs: Extending high-dimensional
  data analysis to networks and other irregular domains 30~(3) (2013) 83--98.

\bibitem{xia2021graph}
F.~Xia, K.~Sun, S.~Yu, A.~Aziz, L.~Wan, S.~Pan, H.~Liu, Graph learning: A
  survey, IEEE Transactions on Artificial Intelligence 2~(2) (2021) 109--127.

\bibitem{battiston2020networks}
F.~Battiston, G.~Cencetti, I.~Iacopini, V.~Latora, M.~Lucas, A.~Patania, J.-G.
  Young, G.~Petri, Networks beyond pairwise interactions: Structure and
  dynamics, Physics reports 874 (2020) 1--92.

\bibitem{krishnagopal2021spectral}
S.~Krishnagopal, G.~Bianconi, Spectral detection of simplicial communities via
  hodge laplacians, Physical Review E 104~(6) (2021) 064303.

\bibitem{battiston2021physics}
F.~Battiston, E.~Amico, A.~Barrat, G.~Bianconi, G.~Ferraz~de Arruda,
  B.~Franceschiello, I.~Iacopini, S.~K{\'e}fi, V.~Latora, Y.~Moreno, et~al.,
  The physics of higher-order interactions in complex systems, Nature Physics
  17~(10) (2021) 1093--1098.

\bibitem{zhang2023higher}
Y.~Zhang, M.~Lucas, F.~Battiston, Higher-order interactions shape collective
  dynamics differently in hypergraphs and simplicial complexes, Nature
  communications 14~(1) (2023) 1605.

\bibitem{iacopini2019simplicial}
I.~Iacopini, G.~Petri, A.~Barrat, V.~Latora, Simplicial models of social
  contagion, Nature communications 10~(1) (2019) 2485.

\bibitem{malizia2024reconstructing}
F.~Malizia, A.~Corso, L.~V. Gambuzza, G.~Russo, V.~Latora, M.~Frasca,
  Reconstructing higher-order interactions in coupled dynamical systems, Nature
  Communications 15~(1) (2024) 5184.

\bibitem{lin2024higher}
Z.~Lin, L.~Han, M.~Feng, Y.~Liu, M.~Tang, Higher-order non-markovian social
  contagions in simplicial complexes, Communications Physics 7~(1) (2024) 175.

\bibitem{schaub2020random}
M.~T. Schaub, A.~R. Benson, P.~Horn, G.~Lippner, A.~Jadbabaie, Random walks on
  simplicial complexes and the normalized hodge 1-laplacian, SIAM Review 62~(2)
  (2020) 353--391.

\bibitem{carlsson2009topology}
G.~Carlsson, Topological methods for the analysis of high dimensional data sets
  and 3d object recognition, Bulletin of the American Mathematical Society
  46~(2) (2009) 255--265.

\bibitem{edelsbrunner2010computational}
H.~Edelsbrunner, J.~Harer, Computational topology: an introduction, American
  Mathematical Society, 2010.

\bibitem{wasserman2018topological}
L.~Wasserman, Topological Data Analysis, Johns Hopkins University Press, 2018.

\bibitem{otter2017roadmap}
N.~Otter, M.~Porter, U.~Tillmann, V.~Vitelli, G.~Carlsson, A roadmap for the
  computation of persistent homology, EPJ Data Science 6~(1) (2017) 17.

\bibitem{ghrist2008barcodes}
R.~Ghrist, Barcodes: the persistent topology of data, Bulletin of the American
  Mathematical Society 45~(1) (2008) 61--75.

\bibitem{edelsbrunner2022computational}
H.~Edelsbrunner, J.~Harer, Computational Topology: An Introduction, American
  Mathematical Society, 2022.

\bibitem{hensel2021survey}
F.~Hensel, M.~Moor, B.~Rieck, A survey of topological machine learning methods,
  Frontiers in Artificial Intelligence 4 (2021) 681108.

\bibitem{bispo2024emergence}
B.~C. Bispo, F.~A. Santos, J.~R. D.~O. Neto, J.~B. Lima, Emergence of
  higher-order functional brain connectivity with hypergraph signal processing,
  in: 2024 32nd European Signal Processing Conference (EUSIPCO), IEEE, 2024,
  pp. 1332--1336.

\bibitem{barbarossa2016introduction}
S.~Barbarossa, M.~Tsitsvero, An introduction to hypergraph signal processing,
  in: {IEEE} Int. Conf. Acoust., Speech and Signal Process., 2016, pp.
  6425--6429.

\bibitem{barbarossa2020topologicalsimplicial}
S.~Barbarossa, S.~Sardellitti, Topological signal processing over simplicial
  complexes, IEEE Transactions on Signal Processing 68 (2020) 2992--3007.

\bibitem{barbarossa2020topological}
S.~Barbarossa, S.~Sardellitti, Topological signal processing: Making sense of
  data building on multiway relations, IEEE Signal Processing Magazine 37~(6)
  (2020) 174--183.

\bibitem{schaub2021signal}
M.~T. Schaub, Y.~Zhu, J.-B. Seby, T.~M. Roddenberry, S.~Segarra, Signal
  processing on higher-order networks: Livin'on the edge... and beyond, Signal
  Processing 187 (2021) 108149.

\bibitem{feng2019hypergraph}
Y.~Feng, H.~You, Z.~Zhang, R.~Ji, Y.~Gao, Hypergraph neural networks, in:
  Proceedings of the AAAI conference on artificial intelligence, Vol.~33, 2019,
  pp. 3558--3565.

\bibitem{antelmi2023survey}
A.~Antelmi, G.~Cordasco, M.~Polato, V.~Scarano, C.~Spagnuolo, D.~Yang, A survey
  on hypergraph representation learning, ACM Computing Surveys 56~(1) (2023)
  1--38.

\bibitem{papillon2023architectures}
M.~Papillon, S.~Sanborn, M.~Hajij, N.~Miolane, Architectures of topological
  deep learning: A survey of message-passing topological neural networks, arXiv
  preprint arXiv:2304.10031 (2023).

\bibitem{papamarkou2024position}
T.~Papamarkou, T.~Birdal, M.~M. Bronstein, G.~E. Carlsson, J.~Curry, Y.~Gao,
  M.~Hajij, R.~Kwitt, P.~Lio, P.~Di~Lorenzo, et~al., Position: Topological deep
  learning is the new frontier for relational learning, in: Forty-first
  International Conference on Machine Learning, 2024.

\bibitem{sanborn2024beyond}
S.~Sanborn, J.~Mathe, M.~Papillon, D.~Buracas, H.~J. Lillemark, C.~Shewmake,
  A.~Bertics, X.~Pennec, N.~Miolane, Beyond euclid: An illustrated guide to
  modern machine learning with geometric, topological, and algebraic
  structures, arXiv preprint arXiv:2407.09468 (2024).

\bibitem{hajij2206topological}
M.~Hajij, G.~Zamzmi, T.~Papamarkou, N.~Miolane, A.~Guzm{\'a}n-S{\'a}enz, K.~N.
  Ramamurthy, T.~Birdal, T.~K. Dey, S.~Mukherjee, S.~N. Samaga, et~al.,
  Topological deep learning: Going beyond graph data, may 2023, arXiv preprint
  arXiv:2206.00606.

\bibitem{ebli2020simplicial}
S.~Ebli, M.~Defferrard, G.~Spreemann, Simplicial neural networks, in: NeurIPS
  workshops, TDA {\&} Beyond, preprint arXiv:2010.03633, 2020.

\bibitem{yang2022simplicialnn}
M.~Yang, E.~Isufi, G.~Leus, Simplicial convolutional neural networks, in:
  ICASSP 2022-2022 IEEE International Conference on Acoustics, Speech and
  Signal Processing (ICASSP), IEEE, 2022, pp. 8847--8851.

\bibitem{battiloro2024generalized}
C.~Battiloro, L.~Testa, L.~Giusti, S.~Sardellitti, P.~Di~Lorenzo,
  S.~Barbarossa, Generalized simplicial attention neural networks, IEEE
  Transactions on Signal and Information Processing over Networks (2024).

\bibitem{goldberg2002combinatorial}
T.~E. Goldberg, Combinatorial laplacians of simplicial complexes, Ph.D. thesis,
  Bard College (2002).

\bibitem{sardellitti2021topological}
S.~Sardellitti, S.~Barbarossa, L.~Testa, Topological signal processing over
  cell complexes, in: 2021 55th Asilomar Conference on Signals, Systems, and
  Computers, IEEE, 2021, pp. 1558--1562.

\bibitem{roddenberry2022signal}
T.~M. Roddenberry, M.~T. Schaub, M.~Hajij, Signal processing on cell complexes,
  in: ICASSP 2022-2022 IEEE International Conference on Acoustics, Speech and
  Signal Processing (ICASSP), IEEE, 2022, pp. 8852--8856.

\bibitem{sardellitti2024topological}
S.~Sardellitti, S.~Barbarossa, Topological signal processing over generalized
  cell complexes, IEEE Transactions on Signal Processing (2024).

\bibitem{yang2022simplicial}
M.~Yang, E.~Isufi, M.~T. Schaub, G.~Leus, Simplicial convolutional filters,
  IEEE Transactions on Signal Processing 70 (2022) 4633--4648.

\bibitem{bianconi2021topological}
G.~Bianconi, The topological dirac equation of networks and simplicial
  complexes, Journal of Physics: Complexity 2~(3) (2021) 035022.

\bibitem{calmon2023dirac}
L.~Calmon, M.~T. Schaub, G.~Bianconi, Dirac signal processing of higher-order
  topological signals, New Journal of Physics 25~(9) (2023) 093013.

\bibitem{grande2024disentangling}
V.~P. Grande, M.~T. Schaub, Disentangling the spectral properties of the hodge
  laplacian: not all small eigenvalues are equal, in: ICASSP 2024-2024 IEEE
  International Conference on Acoustics, Speech and Signal Processing (ICASSP),
  IEEE, 2024, pp. 9896--9900.

\bibitem{rioul1991wavelets}
O.~Rioul, M.~Vetterli, Wavelets and signal processing, IEEE signal processing
  magazine 8~(4) (1991) 14--38.

\bibitem{tovsic2011dictionary}
I.~To{\v{s}}i{\'c}, P.~Frossard, Dictionary learning, IEEE Signal Processing
  Magazine 28~(2) (2011) 27--38.

\bibitem{tsitsvero2015degrees}
M.~Tsitsvero, S.~Barbarossa, On the degrees of freedom of signals on graphs,
  in: 2015 23rd European Signal Processing Conference (EUSIPCO), IEEE, 2015,
  pp. 1506--1510.

\bibitem{roddenberry2022hodgelets}
T.~M. Roddenberry, F.~Frantzen, M.~T. Schaub, S.~Segarra, Hodgelets: Localized
  spectral representations of flows on simplicial complexes, in: ICASSP
  2022-2022 IEEE International Conference on Acoustics, Speech and Signal
  Processing (ICASSP), IEEE, 2022, pp. 5922--5926.

\bibitem{battiloro2023topologicalslepians}
C.~Battiloro, P.~Di~Lorenzo, S.~Barbarossa, Topological slepians: Maximally
  localized representations of signals over simplicial complexes, in: ICASSP
  2023-2023 IEEE International Conference on Acoustics, Speech and Signal
  Processing (ICASSP), IEEE, 2023, pp. 1--5.

\bibitem{saito2023multiscale}
N.~Saito, S.~C. Schonsheck, E.~Shvarts, Multiscale hodge scattering networks
  for data analysis, arXiv preprint arXiv:2311.10270 (2023).

\bibitem{battiloro2023topological}
C.~Battiloro, S.~Sardellitti, S.~Barbarossa, P.~Di~Lorenzo, Topological signal
  processing over weighted simplicial complexes, in: ICASSP 2023-2023 IEEE
  International Conference on Acoustics, Speech and Signal Processing (ICASSP),
  IEEE, 2023, pp. 1--5.

\bibitem{mateos2019connecting}
G.~Mateos, S.~Segarra, A.~G. Marques, A.~Ribeiro, Connecting the dots:
  Identifying network structure via graph signal processing, {IEEE} Signal
  Process. Mag. 36~(3) (2019) 16--43.

\bibitem{hoppe2024representing}
J.~Hoppe, M.~T. Schaub, Representing edge flows on graphs via sparse cell
  complexes, in: Learning on Graphs Conference, PMLR, 2024, pp. 1--1.

\bibitem{gurugubelli2024simplicial}
S.~Gurugubelli, S.~P. Chepuri, Simplicial complex learning from edge flows via
  sparse clique sampling, in: 2024 32nd European Signal Processing Conference
  (EUSIPCO), IEEE, 2024, pp. 2332--2336.

\bibitem{sardellitti2023probabilistic}
S.~Sardellitti, S.~Barbarossa, Probabilistic topological models over simplicial
  complexes, in: 2023 57th Asilomar Conference on Signals, Systems, and
  Computers, IEEE, 2023, pp. 822--826.

\bibitem{friedman2008sparse}
J.~Friedman, T.~Hastie, R.~Tibshirani, Sparse inverse covariance estimation
  with the graphical lasso, Biostatistics 9~(3) (2008) 432--441.

\bibitem{buciulea2024learninga}
A.~Buciulea, E.~Isufi, G.~Leus, A.~G. Marques, Learning the topology of a
  simplicial complex using simplicial signals: A greedy approach, in: 2024 IEEE
  13rd Sensor Array and Multichannel Signal Processing Workshop (SAM), IEEE,
  2024, pp. 1--5.

\bibitem{buciulea2024learning}
A.~Buciulea, E.~Isufi, G.~Leus, A.~G. Marques, Learning graphs and simplicial
  complexes from data, in: ICASSP 2024-2024 IEEE International Conference on
  Acoustics, Speech and Signal Processing (ICASSP), IEEE, 2024, pp. 9861--9865.

\bibitem{wang2022fullh}
H.~Wang, C.~Ma, H.-S. Chen, Y.-C. Lai, H.-F. Zhang, Full reconstruction of
  simplicial complexes from binary contagion and ising data, Nature
  communications 13~(1) (2022) 3043.

\bibitem{battiloro2023latent}
C.~Battiloro, I.~Spinelli, L.~Telyatnikov, M.~Bronstein, S.~Scardapane,
  P.~Di~Lorenzo, From latent graph to latent topology inference: Differentiable
  cell complex module, Proc. of the Twelfth International Conference on
  Learning Representations (ICLR), preprint arXiv:2305.16174 (2023).

\bibitem{isufi2022convolutional}
E.~Isufi, M.~Yang, Convolutional filtering in simplicial complexes, in: ICASSP
  2022-2022 IEEE International Conference on Acoustics, Speech and Signal
  Processing (ICASSP), IEEE, 2022, pp. 5578--5582.

\bibitem{jia2019graph}
J.~Jia, M.~T. Schaub, S.~Segarra, A.~R. Benson, Graph-based semi-supervised \&
  active learning for edge flows, in: Proceedings of the 25th ACM SIGKDD
  international conference on knowledge discovery \& data mining, 2019, pp.
  761--771.

\bibitem{yang2022simplicialtrend}
M.~Yang, E.~Isufi, Simplicial trend filtering, in: 2022 56th Asilomar
  Conference on Signals, Systems, and Computers, IEEE, 2022, pp. 930--934.

\bibitem{liu2023unrolling}
C.~Liu, G.~Leus, E.~Isufi, Unrolling of simplicial elasticnet for edge flow
  signal reconstruction, IEEE Open Journal of Signal Processing (2023).

\bibitem{reddy2024recovery}
S.~Reddy, S.~P. Chepuri, Recovery of signals on a simplicial complex from
  subsampled neighbourhood aggregation, IEEE Signal Processing Letters (2024).

\bibitem{reddy2024sampling}
T.~S. Reddy, S.~P. Chepuri, Sampling and recovery of signals over product cell
  structures, in: ICASSP 2024-2024 IEEE International Conference on Acoustics,
  Speech and Signal Processing (ICASSP), IEEE, 2024, pp. 13191--13195.

\bibitem{gurugubelli2023gaussian}
S.~Gurugubelli, S.~P. Chepuri, Gaussian processes for edge flow prediction with
  active learning, in: 2023 57th Asilomar Conference on Signals, Systems, and
  Computers, IEEE, 2023, pp. 809--813.

\bibitem{bodnar2021weisfeiler}
C.~Bodnar, F.~Frasca, Y.~Wang, N.~Otter, G.~F. Montufar, P.~Lio, M.~Bronstein,
  Weisfeiler and lehman go topological: Message passing simplicial networks,
  in: International Conference on Machine Learning, PMLR, 2021, pp. 1026--1037.

\bibitem{bodnar2021weisfeilerCell}
C.~Bodnar, F.~Frasca, N.~Otter, Y.~Wang, P.~Lio, G.~F. Montufar, M.~Bronstein,
  Weisfeiler and lehman go cellular: Cw networks, Advances in neural
  information processing systems 34 (2021) 2625--2640.

\bibitem{roddenberry_icml_21}
T.~M. Roddenberry, N.~Glaze, S.~Segarra, Principled simplicial neural networks
  for trajectory prediction, Vol. 139, 2021, pp. 9020--9029.

\bibitem{rossi2020sign}
E.~Rossi, F.~Frasca, B.~Chamberlain, D.~Eynard, M.~Bronstein, F.~Monti, Sign:
  Scalable inception graph neural networks, Proc. of ICML 2020 Workshop on
  Graph Representation Learning, preprint arXiv:2004.11198 7 (2020) 15.

\bibitem{gurugubelli2024sann}
S.~Gurugubelli, S.~P. Chepuri, Sann: Simple yet powerful simplicial-aware
  neural networks, in: The Twelfth International Conference on Learning
  Representations, 2024.

\bibitem{gama2020graphs}
F.~Gama, E.~Isufi, G.~Leus, A.~Ribeiro, Graphs, convolutions, and neural
  networks: From graph filters to graph neural networks, IEEE Signal Processing
  Magazine 37~(6) (2020) 128--138.

\bibitem{yang2023convolutional}
M.~Yang, E.~Isufi, Convolutional learning on simplicial complexes, arXiv
  preprint arXiv:2301.11163 (2023).

\bibitem{zhou2024facilitating}
C.~Zhou, X.~Wang, M.~Zhang, Facilitating graph neural networks with random walk
  on simplicial complexes, Advances in Neural Information Processing Systems 36
  (2024).

\bibitem{huang2024higher}
Y.~Huang, Y.~Zeng, Q.~Wu, L.~L{\"u}, Higher-order graph convolutional network
  with flower-petals laplacians on simplicial complexes, in: Proceedings of the
  AAAI Conference on Artificial Intelligence, Vol.~38, 2024, pp. 12653--12661.

\bibitem{chen2022bscnets}
Y.~Chen, Y.~R. Gel, H.~V. Poor, Bscnets: Block simplicial complex neural
  networks, in: Proceedings of the aaai conference on artificial intelligence,
  Vol.~36, 2022, pp. 6333--6341.

\bibitem{yang2022efficient}
R.~Yang, F.~Sala, P.~Bogdan, Efficient representation learning for higher-order
  data with simplicial complexes, in: Learning on Graphs Conference, PMLR,
  2022, pp. 13--1.

\bibitem{Kipf2017-GCN}
T.~N. Kipf, M.~Welling, Semi-supervised classification with graph convolutional
  networks, 2017, pp. 1--14.

\bibitem{yan2025binarized}
Y.~Yan, E.~E. Kuruoglu, Binarized simplicial convolutional neural networks,
  Neural Networks 183 (2025) 106928.

\bibitem{battiloro2024tangent}
C.~Battiloro, Z.~Wang, H.~Riess, P.~Di~Lorenzo, A.~Ribeiro, Tangent bundle
  convolutional learning: from manifolds to cellular sheaves and back, IEEE
  Transactions on Signal Processing (2024).

\bibitem{giusti2022simplicial}
L.~Giusti, C.~Battiloro, P.~Di~Lorenzo, S.~Sardellitti, S.~Barbarossa,
  Simplicial attention neural networks, arXiv preprint arXiv:2203.07485 (2022).

\bibitem{goh2022simplicial}
C.~W.~J. Goh, C.~Bodnar, P.~Lio, Simplicial attention networks, in: ICLR 2022
  Workshop on Geometrical and Topological Representation Learning, arXiv
  preprint arXiv:2204.09455.

\bibitem{hajij2022higher}
M.~Hajij, G.~Zamzmi, T.~Papamarkou, N.~Miolane, A.~Guzm{\'a}n-S{\'a}enz, K.~N.
  Ramamurthy, Higher-order attention networks, arXiv preprint arXiv:2206.00606
  2~(3) (2022) 4.

\bibitem{lee2022sgat}
S.~H. Lee, F.~Ji, W.~P. Tay, Sgat: Simplicial graph attention network, arXiv
  preprint arXiv:2207.11761 (2022).

\bibitem{cinque2023pooling}
D.~M. Cinque, C.~Battiloro, P.~Di~Lorenzo, Pooling strategies for simplicial
  convolutional networks, in: ICASSP 2023-2023 IEEE International Conference on
  Acoustics, Speech and Signal Processing (ICASSP), IEEE, 2023, pp. 1--5.

\bibitem{mcguire2023nervepool}
S.~McGuire, E.~Munch, M.~Hirn, Nervepool: A simplicial pooling layer, In 2023
  Joint Mathematics Meetings (JMM 2023), arXiv preprint arXiv:2305.06315
  (2023).

\bibitem{yang2023hodge}
M.~Yang, V.~Borovitskiy, E.~Isufi, Hodge-compositional edge gaussian processes,
  in: International Conference on Artificial Intelligence and Statistics, PMLR,
  2024, pp. 3754--3762.

\bibitem{alain2023gaussian}
M.~Alain, S.~Takao, B.~Paige, M.~P. Deisenroth, Gaussian processes on cellular
  complexes, Proc. of International Conference on Machine Learning (ICML),
  arXiv preprint arXiv:2311.01198 (2023).

\bibitem{gurugubelli2024gaussian}
S.~Gurugubelli, S.~P. Chepuri, Gaussian processes for predicting simplicial
  closure, in: 2024 IEEE 13rd Sensor Array and Multichannel Signal Processing
  Workshop (SAM), IEEE, 2024, pp. 1--5.

\bibitem{navarro2024sc}
M.~Navarro, S.~Segarra, Sc-mad: Mixtures of higher-order networks for data
  augmentation, in: ICASSP 2024-2024 IEEE International Conference on
  Acoustics, Speech and Signal Processing (ICASSP), IEEE, 2024, pp.
  13446--13450.

\bibitem{zhang2017mixup}
H.~Zhang, mixup: Beyond empirical risk minimization, arXiv preprint
  arXiv:1710.09412 (2017).

\bibitem{madhu2024toposrl}
H.~Madhu, S.~P. Chepuri, Toposrl: topology preserving self-supervised
  simplicial representation learning, Advances in Neural Information Processing
  Systems 36 (2024).

\bibitem{mollers2024hodge}
A.~M{\"o}llers, A.~Immer, V.~Fortuin, E.~Isufi, Hodge-aware contrastive
  learning, in: ICASSP 2024-2024 IEEE International Conference on Acoustics,
  Speech and Signal Processing (ICASSP), IEEE, 2024, pp. 9746--9750.

\bibitem{Volterra2021}
G.~Leus, M.~Yang, M.~Coutino, E.~Isufi, Topological volterra filters, in:
  ICASSP 2021 - 2021 IEEE International Conference on Acoustics, Speech and
  Signal Processing (ICASSP), 2021, pp. 5385--5399.
\newblock \href {https://doi.org/10.1109/ICASSP39728.2021.9414275}
  {\path{doi:10.1109/ICASSP39728.2021.9414275}}.

\bibitem{lutkepohl2005new}
H.~L{\"u}tkepohl, New introduction to multiple time series analysis, Springer
  Science \& Business Media, 2005.

\bibitem{onlzam2019}
B.~{Zaman}, L.~{Ramos}, D.~{Romero}, B.~{Beferull-Lozano}, Online topology
  identification from vector autoregressive time series, IEEE Transactions on
  Signal Processing 69 (2021) 210--225.
\newblock \href {https://doi.org/10.1109/TSP.2020.3042940}
  {\path{doi:10.1109/TSP.2020.3042940}}.

\bibitem{RohanDSLW}
R.~Money, J.~Krishnan, B.~Beferull-Lozano, Online non-linear topology
  identification from graph-connected time series, in: 2021 IEEE Data Science
  and Learning Workshop (DSLW), 2021, pp. 1--6.
\newblock \href {https://doi.org/10.1109/DSLW51110.2021.9523399}
  {\path{doi:10.1109/DSLW51110.2021.9523399}}.

\bibitem{RohanMLSP}
{R. Money}, J.~Krishnan, B.~Beferull-Lozano, Random feature approximation for
  online nonlinear graph topology identification, IEEE MLSP (2021).

\bibitem{RohanTSP}
{{R. Money}}, J.~Krishnan, B.~Beferull-Lozano, Sparse online learning with
  kernels using random features for estimating nonlinear dynamic graphs, IEEE
  Transactions on Signal Processing 71 (2023) 2027--2042.
\newblock \href {https://doi.org/10.1109/TSP.2023.3282068}
  {\path{doi:10.1109/TSP.2023.3282068}}.

\bibitem{nonshe2019}
Y.~Shen, G.~Giannakis, B.~Baingana, Nonlinear structural vector autoregressive
  models with application to directed brain networks, IEEE Transactions on
  Signal Processing 67 (2019) 5325--5339.

\bibitem{veedu2021topology}
M.~Veedu, D.~Harish, M.~Salapaka, Topology learning of linear dynamical systems
  with latent nodes using matrix decomposition, IEEE Transactions on Automatic
  Control (2021).

\bibitem{GVARElvin}
E.~Isufi, A.~Loukas, N.~Perraudin, G.~Leus, Forecasting time series with
  {VARMA} recursions on graphs, IEEE Transactions on Signal Processing 67~(18)
  (2019) 4870--4885.
\newblock \href {https://doi.org/10.1109/TSP.2019.2929930}
  {\path{doi:10.1109/TSP.2019.2929930}}.

\bibitem{Grpah_Filter_Elvin}
E.~Isufi, F.~Gama, D.~Shuman, S.~Segarra, Graph filters for signal processing
  and machine learning on graphs, IEEE Transactions on Signal Processing (2024)
  1--32\href {https://doi.org/10.1109/TSP.2024.3349788}
  {\path{doi:10.1109/TSP.2024.3349788}}.

\bibitem{facotor_model}
C.~Lam, Q.~Yao, {Factor modeling for high-dimensional time series: Inference
  for the number of factors}, The Annals of Statistics 40~(2) (2012) 694 --
  726.

\bibitem{Shrinkage}
N.~Lee, H.~Choi, S.~Kim, Bayes shrinkage estimation for high-dimensional var
  models with scale mixture of normal distributions for noise, Computational
  Statistics and Data Analysis 101 (2016) 250--276.

\bibitem{dim_red}
J.~Chang, B.~Guo, Q.~Yao, Principal component analysis for second-order
  stationary vector time series, The Annals of Statistics 46~(5) (2018)
  2094--2124.

\bibitem{extra3}
L.~Gruber, G.~Kastner, \href{https://arxiv.org/abs/2206.04902}{Forecasting
  macroeconomic data with bayesian vars: Sparse or dense? it depends!} (2023).
\newblock \href {http://arxiv.org/abs/2206.04902} {\path{arXiv:2206.04902}}.
\newline\urlprefix\url{https://arxiv.org/abs/2206.04902}

\bibitem{extra4}
G.~Kastner, F.~Huber,
  \href{https://onlinelibrary.wiley.com/doi/abs/10.1002/for.2680}{Sparse
  bayesian vector autoregressions in huge dimensions}, Journal of Forecasting
  39~(7) (2020) 1142--1165.
\newblock \href {https://doi.org/https://doi.org/10.1002/for.2680}
  {\path{doi:https://doi.org/10.1002/for.2680}}.
\newline\urlprefix\url{https://onlinelibrary.wiley.com/doi/abs/10.1002/for.2680}

\bibitem{extra5}
G.~Primiceri, \href{https://doi.org/10.1111/j.1467-937X.2005.00353.x}{{Time
  Varying Structural Vector Autoregressions and Monetary Policy}}, The Review
  of Economic Studies 72~(3) (2005) 821--852.
\newblock \href {https://doi.org/10.1111/j.1467-937X.2005.00353.x}
  {\path{doi:10.1111/j.1467-937X.2005.00353.x}}.
\newline\urlprefix\url{https://doi.org/10.1111/j.1467-937X.2005.00353.x}

\bibitem{Krishnan_2024}
J.~Krishnan, R.~Money, B.~Beferull-Lozano, E.~Isufi, Simplicial vector
  autoregressive models, IEEE Transactions on Signal Processing (2024)
  1--15,\href {https://doi.org/10.1109/TSP.2024.3503063}
  {\path{doi:10.1109/TSP.2024.3503063}}.

\bibitem{Joshin_SVAR}
J.~Krishnan, R.~Money, B.~Beferull-Lozano, E.~Isufi, Simplicial vector
  autoregressive model for streaming edge flows, in: ICASSP 2023 - 2023 IEEE
  International Conference on Acoustics, Speech and Signal Processing (ICASSP),
  2023, pp. 1--5.
\newblock \href {https://doi.org/10.1109/ICASSP49357.2023.10096095}
  {\path{doi:10.1109/ICASSP49357.2023.10096095}}.

\bibitem{Rohan_backcasting2024}
R.~Money, J.~Krishnan, B.~Beferull-Lozano, E.~Isufi, Evolution backcasting of
  edge flows from partial observations using simplicial vector autoregressive
  models, in: ICASSP 2024 - 2024 IEEE International Conference on Acoustics,
  Speech and Signal Processing (ICASSP), 2024, pp. 9516--9520.
\newblock \href {https://doi.org/10.1109/ICASSP48485.2024.10448180}
  {\path{doi:10.1109/ICASSP48485.2024.10448180}}.

\bibitem{nguyen2024}
D.~T. Nguyen, K.~Slavakis, D.~Pados,
  \href{https://arxiv.org/abs/2409.05135}{Imputation of time-varying edge flows
  in graphs by multilinear kernel regression and manifold learning} (2024).
\newblock \href {http://arxiv.org/abs/2409.05135} {\path{arXiv:2409.05135}}.
\newline\urlprefix\url{https://arxiv.org/abs/2409.05135}

\bibitem{NguyenOJSP2024}
D.~T. Nguyen, K.~Slavakis, Multilinear kernel regression and imputation via
  manifold learning, IEEE Open Journal of Signal Processing 5 (2024)
  1073--1088.
\newblock \href {https://doi.org/10.1109/OJSP.2024.3444707}
  {\path{doi:10.1109/OJSP.2024.3444707}}.

\bibitem{Rohan_SPL23}
R.~Money, J.~Krishnan, B.~Beferull-Lozano, E.~Isufi, Online edge flow
  imputation on networks, IEEE Signal Processing Letters 30 (2023) 115--119.
\newblock \href {https://doi.org/10.1109/LSP.2022.3221846}
  {\path{doi:10.1109/LSP.2022.3221846}}.

\bibitem{RohanAsilomar2024}
R.~Money, M.~Sabbaqi, J.~Krishnan, B.~Beferull-Lonano, E.~Isufi, Kalman
  filtering for simplicial processes, in: Asilomar 2024 - 2024 IEEE
  International Conference, 2024.

\bibitem{chen2022time}
Y.~Chen, Y.~Gel, H.~V. Poor, Time-conditioned dances with simplicial complexes:
  Zigzag filtration curve based supra-hodge convolution networks for
  time-series forecasting, Advances in Neural Information Processing Systems 35
  (2022) 8940--8953.

\bibitem{marinucci2024topological}
L.~Marinucci, C.~Battiloro, P.~Di~Lorenzo, Topological adaptive learning over
  cell complexes, in: 2024 32nd European Signal Processing Conference
  (EUSIPCO), IEEE, 2024, pp. 832--836.

\bibitem{sandryhaila2014big}
A.~Sandryhaila, J.~M. Moura, Big data analysis with signal processing on
  graphs: Representation and processing of massive data sets with irregular
  structure, IEEE signal processing magazine 31~(5) (2014) 80--90.

\bibitem{sabbaqi2023graph}
M.~Sabbaqi, E.~Isufi, Graph-time convolutional neural networks: Architecture
  and theoretical analysis, IEEE Transactions on Pattern Analysis and Machine
  Intelligence (2023).

\bibitem{Roddenberry2022}
T.~M. Roddenberry, M.~T. Schaub, M.~Hajij, Signal processing on cell complexes,
  in: ICASSP 2022 - 2022 IEEE International Conference on Acoustics, Speech and
  Signal Processing (ICASSP), 2022, pp. 8852--8856.
\newblock \href {https://doi.org/10.1109/ICASSP43922.2022.9747233}
  {\path{doi:10.1109/ICASSP43922.2022.9747233}}.

\bibitem{kerimov2024sensor}
B.~Kerimov, V.~Pons, S.~Pritsis, R.~Taormina, F.~Tscheikner-Gratl, Sensor
  placement and state estimation in water distribution systems using edge
  gaussian processes, Engineering Proceedings 69~(1) (2024) 150.

\bibitem{kerimov2024towards}
B.~Kerimov, R.~Taormina, F.~Tscheikner-Gratl, Towards transferable metamodels
  for water distribution systems with edge-based graph neural networks, Water
  Research (2024) 121933.

\bibitem{smith2022physics}
K.~D. Smith, F.~Seccamonte, A.~Swami, F.~Bullo, Physics-informed implicit
  representations of equilibrium network flows, Advances in Neural Information
  Processing Systems 35 (2022) 7211--7221.

\bibitem{chen2023learning}
Y.~Chen, R.~A. Jacob, Y.~R. Gel, J.~Zhang, H.~V. Poor, Learning power grid
  outages with higher-order topological neural networks, IEEE Transactions on
  Power Systems 39~(1) (2023) 720--732.

\bibitem{xia2024locational}
W.~Xia, Y.~Li, L.~Yu, D.~He, Locational detection of false data injection
  attacks in the edge space via hodge graph neural network for smart grids,
  IEEE Transactions on Smart Grid (2024).

\bibitem{liu2024hodge}
C.~Liu, E.~Isufi, Hodge-aware matched subspace detectors, in: 2024 32nd
  European Signal Processing Conference (EUSIPCO), IEEE, 2024, pp. 817--821.

\bibitem{hu2021graph}
W.~Hu, J.~Pang, X.~Liu, D.~Tian, C.-W. Lin, A.~Vetro, Graph signal processing
  for geometric data and beyond: Theory and applications, IEEE Transactions on
  Multimedia 24 (2021) 3961--3977.

\bibitem{taubin1995signal}
G.~Taubin, A signal processing approach to fair surface design, in: Proceedings
  of the 22nd annual conference on Computer graphics and interactive
  techniques, 1995, pp. 351--358.

\bibitem{taubin1996optimal}
G.~Taubin, T.~Zhang, G.~Golub, Optimal surface smoothing as filter design, in:
  Eur. Conf. Comput. Vision, 1996, pp. 283--292.

\bibitem{crane2013digital}
K.~Crane, F.~De~Goes, M.~Desbrun, P.~Schr{\"o}der, Digital geometry processing
  with discrete exterior calculus, in: ACM SIGGRAPH 2013 Courses, 2013, pp.
  1--126.

\bibitem{de2016vector}
F.~De~Goes, M.~Desbrun, Y.~Tong, Vector field processing on triangle meshes,
  in: ACM SIGGRAPH 2016 Courses, 2016, pp. 1--49.

\bibitem{brandt2017spectral}
C.~Brandt, L.~Scandolo, E.~Eisemann, K.~Hildebrandt, Spectral processing of
  tangential vector fields, in: Computer Graphics Forum, Vol.~36, Wiley Online
  Library, 2017, pp. 338--353.

\bibitem{keros2023spectral}
A.~Keros, K.~Subr, Spectral coarsening with hodge laplacians, in: ACM SIGGRAPH
  2023 Conference Proceedings, 2023, pp. 1--11.

\bibitem{smirnov2021hodgenet}
D.~Smirnov, J.~Solomon, Hodgenet: Learning spectral geometry on triangle
  meshes, ACM Transactions on Graphics (TOG) 40~(4) (2021) 1--11.

\bibitem{wiersma2022deltaconv}
R.~Wiersma, A.~Nasikun, E.~Eisemann, K.~Hildebrandt, Deltaconv: anisotropic
  operators for geometric deep learning on point clouds, ACM Transactions on
  Graphics (TOG) 41~(4) (2022) 1--10.

\bibitem{jiang2011statistical}
X.~Jiang, L.-H. Lim, Y.~Yao, Y.~Ye, Statistical ranking and combinatorial hodge
  theory, Mathematical Programming 127~(1) (2011) 203--244.

\bibitem{wei2022hodge}
R.~K.~J. Wei, J.~Wee, V.~E. Laurent, K.~Xia, Hodge theory-based biomolecular
  data analysis, Scientific Reports 12~(1) (2022) 9699.

\bibitem{wieder2020compact}
O.~Wieder, S.~Kohlbacher, M.~Kuenemann, A.~Garon, P.~Ducrot, T.~Seidel,
  T.~Langer, A compact review of molecular property prediction with graph
  neural networks, Drug Discovery Today: Technologies 37 (2020) 1--12.

\bibitem{zhang2023survey}
M.~Zhang, M.~Qamar, T.~Kang, Y.~Jung, C.~Zhang, S.-H. Bae, C.~Zhang, A survey
  on graph diffusion models: Generative ai in science for molecule, protein and
  material, arXiv preprint arXiv:2304.01565 (2023).

\bibitem{jiang2021topological}
Y.~Jiang, D.~Chen, X.~Chen, T.~Li, G.-W. Wei, F.~Pan, Topological
  representations of crystalline compounds for the machine-learning prediction
  of materials properties, npj computational materials 7~(1) (2021) 28.

\bibitem{bernardez2024icml}
G.~Bern{\'a}rdez, L.~Telyatnikov, M.~Montagna, F.~Baccini, M.~Papillon,
  M.~Ferriol-Galm{\'e}s, M.~Hajij, T.~Papamarkou, M.~S. Bucarelli, O.~Zaghen,
  et~al., Icml topological deep learning challenge 2024: Beyond the graph
  domain, arXiv preprint arXiv:2409.05211 (2024).

\bibitem{lee2019coidentification}
H.~Lee, M.~K. Chung, H.~Kang, H.~Choi, S.~Ha, Y.~Huh, E.~Kim, D.~S. Lee,
  Coidentification of group-level hole structures in brain networks via hodge
  laplacian, in: Medical Image Computing and Computer Assisted
  Intervention--MICCAI 2019: 22nd International Conference, Shenzhen, China,
  October 13--17, 2019, Proceedings, Part IV 22, Springer, 2019, pp. 674--682.

\bibitem{anand2023hodge}
D.~V. Anand, M.~K. Chung, Hodge laplacian of brain networks, IEEE transactions
  on medical imaging 42~(5) (2023) 1563--1573.

\bibitem{dakurah2022modelling}
S.~Dakurah, D.~V. Anand, Z.~Chen, M.~K. Chung, Modelling cycles in brain
  networks with the hodge laplacian, in: International Conference on Medical
  Image Computing and Computer-Assisted Intervention, Springer, 2022, pp.
  326--335.

\bibitem{anand2024hodge}
D.~V. Anand, M.~K. Chung, Hodge-decomposition of brain networks, in: 2024 IEEE
  International Symposium on Biomedical Imaging (ISBI), IEEE, 2024, pp. 1--5.

\bibitem{nasrin2019bayesian}
F.~Nasrin, C.~Oballe, D.~Boothe, V.~Maroulas, Bayesian topological learning for
  brain state classification, in: 2019 18th IEEE International Conference On
  Machine Learning And Applications (ICMLA), IEEE, 2019, pp. 1247--1252.

\bibitem{huang2023heterogeneous}
J.~Huang, M.~K. Chung, A.~Qiu, Heterogeneous graph convolutional neural network
  via hodge-laplacian for brain functional data, in: International Conference
  on Information Processing in Medical Imaging, Springer, 2023, pp. 278--290.

\bibitem{park2023convolving}
J.~Park, Y.~Hwang, M.~Kim, M.~K. Chung, G.~Wu, W.~H. Kim, Convolving directed
  graph edges via hodge laplacian for brain network analysis, in: International
  Conference on Medical Image Computing and Computer-Assisted Intervention,
  Springer, 2023, pp. 789--799.

\bibitem{hwang2024multi}
Y.~Hwang, S.~Hwang, G.~Wu, W.~H. Kim, Multi-order simplex-based graph neural
  network for brain network analysis, in: International Conference on Medical
  Image Computing and Computer-Assisted Intervention, Springer, 2024, pp.
  532--541.

\bibitem{pastor2015epidemic}
R.~Pastor-Satorras, C.~Castellano, P.~Van~Mieghem, A.~Vespignani, Epidemic
  processes in complex networks, Reviews of modern physics 87~(3) (2015)
  925--979.

\bibitem{yan2017graph}
X.~Yan, B.~M. Sadler, R.~J. Drost, L.~Y. Paul, K.~Lerman, Graph filters and the
  z-laplacian, IEEE Journal of Selected Topics in Signal Processing 11~(6)
  (2017) 774--784.

\bibitem{tomy2022estimating}
A.~Tomy, M.~Razzanelli, F.~Di~Lauro, D.~Rus, C.~Della~Santina, Estimating the
  state of epidemics spreading with graph neural networks, Nonlinear Dynamics
  109~(1) (2022) 249--263.

\bibitem{liu2024review}
Z.~Liu, G.~Wan, B.~A. Prakash, M.~S. Lau, W.~Jin, A review of graph neural
  networks in epidemic modeling, in: Proceedings of the 30th ACM SIGKDD
  Conference on Knowledge Discovery and Data Mining, 2024, pp. 6577--6587.

\bibitem{barbarossa2023semantic}
S.~Barbarossa, D.~Comminiello, E.~Grassucci, F.~Pezone, S.~Sardellitti,
  P.~Di~Lorenzo, Semantic communications based on adaptive generative models
  and information bottleneck, IEEE Communications Magazine 61~(11) (2023)
  36--41.

\bibitem{strinati2024goal}
E.~C. Strinati, P.~Di~Lorenzo, V.~Sciancalepore, A.~Aijaz, M.~Kountouris,
  D.~G{\"u}nd{\"u}z, P.~Popovski, M.~Sana, P.~A. Stavrou, B.~Soret, et~al.,
  Goal-oriented and semantic communication in {6G} ai-native networks: The
  {6G}-goals approach, Proc. of EuCNC 2024, arXiv preprint arXiv:2402.07573
  (2024).

\bibitem{baccini2022weighted}
F.~Baccini, F.~Geraci, G.~Bianconi, Weighted simplicial complexes and their
  representation power of higher-order network data and topology, Physical
  Review E 106~(3) (2022) 034319.

\bibitem{riihimaki2023simplicial}
H.~Riihim{\"a}ki, Simplicial-connectivity of directed graphs with applications
  to network analysis, SIAM Journal on Mathematics of Data Science 5~(3) (2023)
  800--828.

\bibitem{gong2024higher}
X.~Gong, D.~J. Higham, K.~Zygalakis, G.~Bianconi, Higher-order connection
  laplacians for directed simplicial complexes, Journal of Physics: Complexity
  5~(1) (2024) 015022.

\bibitem{lecha2024higher}
M.~Lecha, A.~Cavallo, F.~Dominici, E.~Isufi, C.~Battiloro, Higher-order
  topological directionality and directed simplicial neural networks, arXiv
  preprint arXiv:2409.08389 (2024).

\end{thebibliography}




\end{document}